\documentclass[11pt,a4paper]{article}
\pdfoutput=1
\usepackage{jheppub}
\usepackage{cleveref}
\usepackage{verbatim}
\usepackage{mathrsfs}
\usepackage{appendix}
\usepackage{mathtools}
\usepackage{float}
\usepackage{subcaption}
\usepackage{alltt}
\usepackage{graphicx}
\usepackage{amssymb,amsmath,amsfonts,mathrsfs,enumerate,latexsym,wasysym}
\usepackage{braket}
\usepackage{xcolor}  
\usepackage{slashed}    
\usepackage{url}
\usepackage{hyperref}
\usepackage{mathtools}
\usepackage{footnotehyper}

\title{\bf  Time evolution of spread complexity in quenched Lipkin-Meshkov-Glick model}

\author{\sf Mir Afrasiar, Jaydeep Kumar Basak, Bidyut Dey, Kunal Pal, Kuntal Pal\footnote{Corresponding author} }

\emailAdd{afrasiar@iitk.ac.in, jaydeep@iitk.ac.in, bidyutd@iitk.ac.in, kunalpal@iitk.ac.in, kuntal@iitk.ac.in}

\affiliation{Department of Physics, 
	Indian Institute of Technology - Kanpur, 
	Kanpur 208016, India}

\abstract{ We use the spread complexity of a time evolved state after a sudden quantum quench in the Lipkin-Meshkov-Glick (LMG) model prepared in the ground state as a probe of quantum phase transition when the system is quenched towards the critical point. By studying the growth of the effective number of elements of the Krylov basis, those contribute to the spread complexity more than a preassigned cut off, we show how the two phases of the LMG model can be distinguished. We also explore the time evolution of spread entropy after both non-critical and critical quenches. We show that the sum contributing to the spread entropy converges slowly in the symmetric phase of the LMG model compared to that of the broken phase, and  for a critical quench, the spread entropy diverges logarithmically at late times. 
}

\begin{document}
\maketitle

\section{Introduction}\label{intro}
Understanding the notion of complexity in quantum systems and field theory has recently gained a lot of attention. One of the reasons behind this surge of interest is the notion that this quantity can be 
used as a probe of black hole interiors, and this has been conjectured to be  given in terms of either 
a suitable  volume or
gravitational action \cite{Susskind:2014rva}. From the viewpoint of the gauge-gravity duality, the important question 
to ask is, what observable would be dual to the complexity of a black hole in terms of a well 
defined quantity on the field theory 
side. On the other hand, the notion of circuit complexity (CC) in quantum systems that is extensively used in computer science is   defined as the difficulty of preparing  a `target state' starting from a particular `reference state' using a given 
set of unitary operators as basic building blocks. Among the most used formulations of the above statement, two of the popular notions are the Nielsen complexity (NC) and the Fubini-Study complexity (FSC). Introduced in \cite{Nielsen1}, 
the NC measures the minimum of a particular cost functional to implement a given task, 
which can be expressed as the geodesic distance in a Riemannian manifold constructed from the unitary operator 
connecting the reference and the target state. Originally defined in terms of the discrete gate sets, 
a notion of NC for field theory was given in \cite{Jefferson:2017sdb} which was further elaborated in the works of \cite{Khan:2018rzm,Bhattacharyya:2018bbv}.  A slightly different notion of complexity is to measure the geodesics distance on the Riemannian parameter manifold of a Hamiltonian, which is induced with the Fubini-Study metric - 
subsequently called as the FSC \cite{Chapman:2017rqy}. 
Related to these  two popular ways of calculating  the complexity, there are also related avenues that have been
explored recently to quantify circuit complexity. For example, several works have appeared related to
the bi-invariant complexity (which is particularly important 
for studying complexity in QFTs) \cite{Yang:2018cgx}, complexity from covariance matrix
\cite{Hackl:2018ptj,Chapman:2018hou}, complexity from the information geometry \cite{DiGiulio:2021oal}, path integral approach to circuit optimization \cite{Bhattacharyya:2018wym}, possible extensions to conformal field theories \cite{Erdmenger:2020sup,  Basteiro:2021ene}, in effective field theory setup \cite{Adhikari:2021ckk} (see the recent review \cite{Chapman:2021jbh} and references therein). 
In this work we  explore a relatively new notion of complexity, namely the spread complexity (SC)
of a time evolved target state, when the parameters of a quantum many-body system showing quantum phase transition is 
suddenly quenched to  a new set of values.

The above mentioned measures of  complexity have also found important applications
 in the context of quantum many body physics.
The three broad sub-areas where the CC have been used recently are (1) studying quantum phase transitions (QPTs), (2) probing non-equilibrium dynamics, typically following a quantum quench and (3) and as an indicator of quantum chaos. Starting from the work of \cite{Liu:2019aji}, which shows nonanalytical behaviors of the NC at the critical points for topological phase transitions in the Kitaev chain, there are quite a few works in this direction to study zero temperature QPT  \cite{sachdev}. The motivation behind these kinds of studies is to notice that an observable will fail to be analytic with a  continuous change of the parameters characterizing the Hamiltonian. For example, the authors of \cite{Xiong:2019xoh} used NC as a probe of topological phase transition in one and two dimensions. 

The work of \cite{Jaiswal:2020nzq} elaborates the use of both NC and FSC to detect ground state QPT in transverse field Ising models. Evolution of NC for for quench dynamics in transverse field Ising model was studied in \cite{Jaiswal:2021tnt} and for  periodic driving in \cite{Camilo:2020gdf}.

The Behaviors of CC were explored for the Bose-Hubbard model in \cite{Sood:2021cuz, Huang:2021xjl} and for the infinite range Lipkin-Meshkov-Glick (LMG) model in \cite{Pal:2022ptv}. On the other hand, the time evolution of CC under a quantum quench and its comparison with entanglement measures of a system can reveal a great deal of information about the underlying dynamics, depending on the integrability property of the Hamiltonian. Works in these directions include CC evolution under smooth mass quench in free field theory \cite{Alves:2018qfv}, scaling of complexity in different quench regions \cite{Camargo:2018eof}, distinctions between different proposals for complexity following a quench \cite{Ali:2018fcz} and multiple quenches \cite{Pal:2022rqq,Gautam:2022gci}. Time evolution of CC involving a classically chaotic Hamiltonian can be used as an effective probe for still elusive quantum signatures of chaos as demonstrated in \cite{Ali:2019zcj, Qu:2021ius}.

One of the still ambiguous features of Nielsen type complexities is the choice of the so called `cost functional' associated with a particular path in the unitary manifold. In the original formulation of Nielsen in the context of quantum computation, the choice of the gate sets and their associated paths are in one's hand, and the cost selection is somewhat natural. However, this statement is not appropriate for the case of real life quantum systems and field theories \cite{Yang:2018nda, Bueno:2019ajd}. 
Consequently, there is still scope for refining the exact nature of quantum state complexity, which is dual to  holographic complexity. In view of the above discussions, a new measure of complexity named the SC for quantum states has gained significant attention. Introduced originally in the context of operator growth in quantum systems, the `Krylov complexity' (KC) measures how an initial simple operator becomes complex under the evolution of a chaotic Hamiltonian \cite{Parker:2018yvk}. The key aspect of calculating the KC includes finding the Krylov basis, an orthonormal basis in the operator Hilbert space. The canonical way for constructing the same is to implement the so called `Lanczos algorithm' \cite{VM}. Roughly the Lanczos algorithm is an iterative procedure that gives two sets of constants $a_{n}$ and $b_{n}$ as output with the auto-correlation function, the inner overlap of a time evolved state at an arbitrary time with the state at the initial time, provided as an input. It was conjectured that the linear growth of the quantity  $b_{n}$ implies a chaotic nature of the evolution \cite{Parker:2018yvk}. Various aspects of this conjecture and possible way of finding the Lanczos coefficients by means of analytical and numerical tools were elaborated in \cite{Barbon:2019wsy,Dymarsky:2019elm,Dymarsky:2021bjq,Zotos:2021uyn,Yates:2021asz,Caputa:2021sib,Kim:2021okd,Caputa:2021ori,Patramanis:2021lkx,Trigueros:2021rwj,Rabinovici:2021qqt,Fan:2022xaa,Heveling:2022hth,Bhattacharjee:2022vlt,Adhikari:2022oxr,Adhikari:2022whf,Muck:2022xfc,Banerjee:2022ime,Carabba:2022itd,Bhattacharya:2022gbz,Rabinovici:2022beu,Liu:2022god}. In a slightly different context, the approach to defining the Krylov subspace and corresponding complexity directly for the unitary evolution of an initial state under a Hamiltonian was introduced in \cite{Balasubramanian:2022tpr}. It was shown in this reference that
the spread of a wave function over all choices of the basis on the Hilbert space is minimized when 
the orthonormal basis is the Krylov basis constructed from the reference state as the starting point. 
In this case, the complete sets of the Lanczos coefficients can be obtained from the `survival amplitude', which is the overlap between the initial reference state and the unitary evolved target state. This notion of SC was explored further in \cite{Caputa:2022eye}  to distinguish different topological phases of the Su-Schrieffer-Heeger (SSH) model and in Kitaev chain \cite{Caputa:2022yju}. It was concluded that the SC shows markedly different behavior in two phases, in contrast to that of the NC. The authors of \cite{Bhattacharjee:2022qjw} used SC of time evolved states to probe weak ergodicity-breaking in the scar states - the states that weakly violate the eigenstate thermalization
hypothesis. 

A generic quantum many body system can show signatures of equilibrium quantum phase transition even when the system is kicked out of equilibrium by employing a sudden change in the parameters of the Hamiltonian. Various information theoretic measures like time dependent fidelity, average work, and irreversible work have been used to locate static critical points.
In this article, our interest is to study the equilibrium signatures of quantum phase transition when a many body system 
having infinite range interaction is driven towards its quantum critical point, by using the SC of the time evolved state as a possible probe. To this end we use the LMG model as a prototype  quantum system having 
infinite range interaction between its constituting particles. Though, due to the long-range nature of the interaction, it cannot be mapped to free fermion models, in the thermodynamic limit the phases of this model can be worked out, and have been studied extensively in the literature \cite{PhysRevE.78.021106,PhysRevLett.99.050402}. Besides these works on the 
static properties of the LMG model, the time-dependent fidelity in a quench scenario was studied \cite{Campbell},
the Nielsen and Fubini-Study complexity in \cite{Pal:2022ptv, Pal:2021jtm} - where all these quantities have been
shown to have some particular behavior as the system moves towards the critical point.  

In this paper, we first construct the Krylov basis, by implementing the Lanczos algorithm starting from 
the pre-quenched LMG model ground state in either of the two characteristics phases of the system. We then obtain  the 
Lanczos coefficients  numerically  \cite{VM}, and also provide a compact form of these coefficients, 
which are then used to generate an analytical expression for the time evolution of the  SC
of the time evolved state after a sudden quench of the parameters of the LMG model.
The SC shows complete revivals while the target state is away from the critical point, and
the magnitudes of these oscillations grow as the system is moved towards the critical point.
To systematically characterize this  growth,  we define a new quantity $N_{eff}$, which we call the effective
number of elements of the Krylov basis which contribute to the growth of the SC, up to 
a predefined cutoff, and it is shown that this 
quantity grows as the critical point is approached.  Furthermore, it is shown that the growth of $N_{eff}$
is different on the two sides of the critical point, thereby providing a way to distinguish between them
in a quench process. When the quench is  a
critical  quench we show that the SC grows quadratically with time, and hence, diverges  late times. 
 
Furthermore we also study  the analogue of K-entropy, defined as spread entropy, in the context of Hamiltonian evolution. The spread entropy shows oscillatory behavior for noncritical quench, while for the critical quench, it shows logarithmic divergence at late times.

The paper is structured as follows. In \cref{SC} we review the basic structure of Krylov complexity and the Lanczos algorithm. 
In \cref{SC_LMG} we study the time evolution of SC under sudden quench in a harmonic oscillator model 
related to the LMG model at the thermodynamic limit. 
We start with a concise review of the thermodynamic limit of the LMG model in \cref{Therm_LMG}. 
Subsequently, we analyze the time evolution of the oscillator ground state under the quenched Hamiltonian and compute the Lanczos coefficients, as well as the SC in \cref{quench} and \cref{Lanczos} respectively. 
In \cref{critical_quench} we conclude the study of SC by analyzing its behavior under a critical quench. 
In \cref{SE}  we study  the time evolution of a related quantity namely, the spread entropy, under critical as well as 
non critical quenches. The time evolution of SC in both the phases of a quenched LMG model is studied in \cref{SC_LMG_Exact}
and its relation with the oscillator model is discussed.
Finally, we summarize our findings in \cref{conclu}. The paper also contains an appendix where we briefly 
analyze the SC in the  SSH model when the initial state is not the ground state of SSH Hamiltonian.

\section{Spread complexity over Krylov  basis}\label{SC}
\label{procedure}
The circuit complexity to prepare a quantum target state $\ket{\psi_{T}}$ starting from a reference state $\ket{\psi_{R}}$ where the latter is related to the former by a unitary transformation, is defined as the minimum number of fundamental gates required in the process. There exist various approaches in the literature for calculating the complexity of a given quantum state transformation. One of the most popular approaches is Nielsen's geometric formulation, where the circuit complexity is given by the minimum value of a particular cost functional assigned to each path on a Riemannian manifold obtained from the unitary transformation connecting the reference and the target states. This procedure essentially reduces to calculating the geodesic distance between two points on the 
unitary manifold representing the target and the reference states, respectively.  

On the other hand, a somewhat more direct approach which is motivated by the notion of Krylov complexity of operator growth in quantum chaotic systems is the recently introduced notion of SC 
over the basis in the Hilbert space of a particular quantum system. To elaborate, let us consider the  
evolution of a state under a `protocol Hamiltonian' $H$, as 
\begin{equation}\label{unitary}
	\ket{\psi(s)}=e^{-iHs}\ket{\psi(s=0)}.
\end{equation}
Here $s$ is an arbitrary parameter characterizing the reference state ($s=0$) and the target state ($s=1$). For time evolutions generated by a system's time-independent quantum mechanical Hamiltonian, this parameter is the time (t). The basic goal of defining the SC is to precisely quantify how the reference state $\ket{\psi_{R}}$ spreads over the Hilbert space. To this end, one first defines a measure of this spreading i.e., a cost function as
\begin{equation}\label{cost}
	\mathcal{C}_{B}(s)=\sum_{n}n|\braket{\psi(s)|B_{n}}|^2~,
\end{equation}
with respect to  some particular complete orthonormal  basis $B$ $\{\ket{B_{n}},n=0, 1, 2, \cdots \}$. Here, the cost function is defined in such a way that it increases if a given wavefunction spreads deeper in the basis $B$. The minimization of this cost function over all the possible choices of the bases defines the SC.

Among all the choices of the bases $B$, a particular basis known as the Krylov basis plays the most significant role in defining the SC. The Krylov basis ($\ket{K_{n}}, n=0, 1, 2 \cdots $), as an orthonormal set of base kets, can be obtained using the Gram-Schmidt orthogonalization procedure on the basis kets of the expansion of eq. (\ref{unitary}). This particular basis is convenient for studying the spread of the initial state over the full Hilbert space.

The key idea behind the construction of the Krylov basis is to write the Hamiltonian in the tri-diagonal basis in the Lanczos algorithm. In this procedure a new basis is defined in terms of the old basis as follows, 
\begin{equation}\label{Krylov-basis}
	\ket{K_{n+1}}=\frac{1}{b_{n+1}}\Big[(H-a_{n})\ket{K_n}-b_{n}\ket{K_{n-1}}\Big]~.
\end{equation}
Here $\ket{K_{0}}=\ket{\psi(0)}$ implies that the algorithm starts with the reference state. The computation of the coefficients $a_{n}, b_{n}$, known as the Lanczos coefficients, plays a pivotal role in implementing the Lanczos algorithm. This information about the Lanczos coefficients is also encoded in the so called `return-amplitude', which is defined as the overlap between the state at any particular value of the circuit parameter $s$ and the initial state, i.e
\begin{equation}
	\mathcal{S}(s)=\braket{\psi(s)|\psi(0)}~.
\end{equation}
The return amplitude is the analogue of the auto-correlation function in the Liouvillian recursion. The expansion of the target state in terms of the Krylov basis is given by
\begin{equation}
\ket{\psi(s)}=\sum_{n}\phi_{n}(s)\ket{K_{n}}~,
\end{equation}
where, the expansion coefficients $\phi_{n}(s)$ satisfy the discrete Schrödinger equation 
\begin{equation}\label{dse}
	i\partial_{s}\phi_{n}(s)=a_{n}\phi_{n}(s)+b_{n}\phi_{n-1}(s)+b_{n+1}\phi_{n+1}(s)~.
\end{equation}  
These coefficients $\phi_n$s can be interpreted as the probability of finding the time-evolved state in the 
$n$th Krylov basis.

The fact that the Krylov basis defined above minimizes the cost function in eq. (\ref{cost}) has been proved recently in \cite{Balasubramanian:2022tpr}. In the Krylov basis, the  expression of the SC becomes particularly simple and given by
\begin{equation}\label{spread_complexity}
	\mathcal{C}(s)=\sum_{n}n|\phi_{n}(s)|^2~~.
\end{equation}

The next two important steps in the  procedure of evaluating the SC consists of determining the Lanczos coefficients from the return amplitude, and finding the $\phi_{n}(s)$ from the eq. (\ref{dse}). For calculating the Lanczos coefficients from the return amplitude, we first need to find the even and odd moments from the expansion as,
\begin{equation}\label{moments}
	\mathcal{S}(s)=\sum_{n}^{\infty}M_{n}\frac{s^n}{n!}~.
\end{equation}
Once we know the full sets of moments, we can extract $a_{n}$s and $b_{n}$s using the standard recursion methods available in the literature \cite{VM} which we briefly review below.  To construct the full set of orthonormal Krylov basis on the Hilbert space, we  start from the given state $\ket{\psi(0)}$, i.e. this is the first of the Krylov state  $\ket{K_{0}}=\ket{\psi(0)}$. Then the recursion relation eq. (\ref{Krylov-basis}) implies that the next basis is  $\ket{K_{1}}=\frac{1}{b_{1}}\Big[(H-a_{0})\ket{K_{0}}\Big]$. Here we have used the fact that $b_{0}=0$. The condition that this state $\ket{K_{1}}$ is orthogonal to the previous state $\ket{K_{0}}$ fixes the unknown coefficient $a_{0}$ to be equal to $\braket{K_{0}|H|K_{0}}$. And the other coefficient $b_{1}$ ensures the normalization of this state. We  continue this recursive process to construct the full set of basis and the general coefficients are  given as 
 \begin{equation}
 	 a_{n}=\braket{K_{n}|H|K_{n}}~,
 \end{equation}
while $b_{n}$s fix the normalization at each step. However in practice, where the above process does not terminate after first few steps, it is more useful to implement the Lanczos algorithm by means of two sets of auxiliary matrices $L_{k}^{(n)}$ and $M_{k}^{(n)}$ constructed from the moments $M_{n}$s of the return amplitude defined in eq. (\ref{moments}). The recursion relations then can be written down in terms of those  $L_{k}^{(n)}$s and $M_{k}^{(n)}$s and finally the Lanczos coefficients can be obtained as $b_{n}=\sqrt{M_{n}^{(n)}}$ and $a_{n}=-L_{n}^{(n)}$ with the initial conditions properly chosen \cite{VM}. 

The authors in \cite{Caputa:2021sib} advanced another elegant method to obtain the Lanczos coefficients for systems where the Hamiltonian (or the Liouvillian) satisfies some particular algebraic relations. In this picture the final state after evolution is noting but a generalised coherent state produced by acting the displacement operator of the associated algebra on the initial state. Consequently, the $a_{n}$s and $b_{n}$s can be computed directly in a simple manner by utilising the `ladder operators' of the algebra. Also it is possible to find $\phi_{n}$s from the expansion of the coherent states in terms of the basis vectors of the Hilbert space. This approach not only provides a straightforward way to construct the Krylov basis but also explains the geometric meaning of the Krylov complexity in terms of the volumes corresponding to the associated information geometry (for details see \cite{Caputa:2021sib}).

After finding $a_{n}$s and $b_{n}$s by using any of the above mentioned methods we can solve the discrete Schrodinger equation (\ref{dse}), for each value of $n$  and obtain $\phi_{n}$s with the initial condition $\phi_{n}(0)=\delta_{n,0}$. We use this procedure to study the time evolution of the SC in sudden quenches of some known quantum systems.

\textbf{Effective number of elements of the Krylov basis.}
Before going to the description of the quantum systems considered in this paper, here we introduce a new quantity which will be used later on to distinguish different phases of a quantum many-body system. For the harmonic oscillator (HO) and the LMG model considered in the main text, we shall see that the ratio of modulus square of  successive coefficients $\phi_{n}$ are smaller than unity, so that the summation in the expression for the SC is convergent, even when there are infinite numbers of Krylov basis. Therefore, the higher elements of the Krylov basis will contribute lesser and lesser amount towards the total summation (see the discussion in subsection 3.3 for a mathematical quantification of these statements). In those cases, to track the spread of the time-evolved state in the Krylov basis, we  introduce a new quantity, which we name as the ``effective number of elements'' $n=N_{eff}$, of the Krylov basis through the following relation (applied at a time instant $t=t_e$) 
\begin{equation}\label{N_eff}
	\mathcal{C}^{(N_{eff}+1)}(t)-\mathcal{C}^{(N_{eff})}(t) 
	=\sum_{n=0}^{N_{eff}+1}n|\phi_{n}(t)|^2 - \sum_{n=0}^{N_{eff}}n|\phi_{n}(t)|^2  \leq \epsilon~,
\end{equation}
is satisfied. Here $\epsilon$ is a  small quantity, which is set to be $\epsilon=0.001$ in the numerical estimations in the next sections, and $t$ denotes the time after a sudden quench. Since the complexity $\mathcal{C}(t)$ is time-dependent, to compute $N_{eff}$ we actually need to apply the above inequality  at some particular  instant of time, which we call $t_e$.

As we shall see in the context of the auxiliary HO and the infinite range LMG model considered 
in sections \ref{SC_LMG} and \ref{SC_LMG_Exact} respectively, this can be used as an effective marker of the QPT 
in these models through the different  scaling relations obeyed by $N_{eff}$ in different phases.

\textbf{Spread entropy:} As we have mentioned before, since $\phi_n$ is the probability of obtaining the time-evolved
state in the $n$th Krylov basis,  one can define an entropy function through the following relation  \cite{Barbon:2019wsy}
\begin{equation}\label{entropy}
S_{K} = -\sum_{n} |\phi_{n}(t)|^2 \log |\phi_{n}(t)|^2~.
\end{equation} 

	The complex conjugate of the auto-correlation function (or the first coefficient $\phi_0$) measures only the overlap 
	of the initial state with the time-evolved state, while the higher coefficients $\phi_n$, has information about the 
	participation of other Krylov basis elements in the time-evolved state.  Therefore, the spread entropy, as defined above,  is a very useful quantity for  measuring delocalisation of 
	the initial wavefunction in the Krylov basis. 
	Similar quantities, computed in terms of the  eigenstate of the pre-quenched Hamiltonian,  have also been studied 
	in the context of the   quenches in interacting quantum many-body systems \cite{torres2014general}.
	We study the time evolution of this quantity after a sudden quench in an HO model related to the LMG model (see section \ref{SE}).

\section{Spread complexity in instantaneous quantum quenches of an oscillator  related to the LMG model}\label{SC_LMG} The time evolution of the SC in a quantum system after a sudden quench can be computed following a few simple steps. For the many-body systems we consider a sudden quench protocol is defined by
changing the system parameters  instantaneously to a new set of values.
We assume that the pre-quenched system is prepared in the lowest state of the Krylov basis, i.e. $\ket{K_{0}}=\ket{\psi(t=0)}=\ket{\psi_i}$, which is not necessarily  the ground state of the system Hamiltonian $H_i$. At $t=0$, we perform the sudden quench changing the system Hamiltonian $H_f$, which subsequently drives the time evolution. The state at any arbitrary time $t$ after the quench can be expressed as 
\begin{equation}\label{psi_ft}
	\ket{\Psi_f(t)}=e^{-iH_f t} \ket{\psi(t=0)} = e^{-iH_f t} \ket{K_{0}} ~.
\end{equation}
Comparing the above equation with \cref{unitary}, one can assume that the real time $t$ plays the role of circuit time $s$ explained in \cref{SC}. Hence the return amplitude, containing all the information about the Lanczos coefficients, is expressed by the following inner product
\begin{equation}\label{SSt}
	\mathcal{S}(t)=\braket{\Psi_f(t) | \psi(t=0)} = \bra{\psi(t=0)}e^{iH_f t}
	\ket{\psi(t=0)}~.
\end{equation}
Utilizing \cref{SSt}, we compute all the Lanczos coefficients which are used to find out the $\phi_{n}(t)$s by solving the discrete Schrödinger equation (\ref{dse}). Following \cref{spread_complexity}, we determine the SC as a function of time using the  Lanczos coefficients mentioned before.

In this article, we mainly focus on the LMG model  of nuclear physics \cite{LIPKIN1965188},
which is one of the most studied example of many-body system involving infinite range interactions. This model also shows a quantum phase transition at critical values of the parameters involved. Considering appropriate Bogoliubov transformations, the LMG model can be diagonalized  in the thermodynamic limit. These mathematical tricks make the computations of the SC simpler, as we show in sequel. 

In \cref{SSH} we briefly consider the time evolution of the SC after a quantum quench in the Su-Schrieffer-Heeger (SSH) model to illustrate the difference between the results in the SSH and the LMG model.


\subsection{Thermodynamic limit of  the LMG model and a related harmonic oscillator model}\label{Therm_LMG}
The LMG  model describes $N$ spin 1/2 self interacting particles  acted upon by an external field. The Hamiltonian of this model can be written in terms of  components $J_{\alpha}$ (with $\alpha=\{x,y,z\}$) of a collective spin operator $J$  as 
\begin{equation}
	\label{LMGHamiltonian}
	\mathcal{H} =-\frac{2}{N}\left(J^{2}_{x}+
	\gamma J^{2}_{y} \right)-2 hJ_{z}~,
\end{equation}
where we have neglected an irrelevent constant energy shift \cite{PhysRevB.71.224420}. In the above equation, $h$ is an externally applied magnetic field, which for convenience we assume to be along the
$z$ direction. The anisotropy of the spin-spin interaction is characterized by the constant
$\gamma$ which varies in the range $0\leq\gamma\leq1$.

The ground state of this model exhibits a second order QPT in the thermodynamic limit $N \rightarrow \infty$ of this Hamiltonian when the value of the external field approaches a critical value $h_c \rightarrow  1$  \cite{PhysRevLett.99.050402, PhysRevE.78.021106}. In this paper, all the results are obtained considering the thermodynamic limit of the LMG model. Furthermore, assuming that the external field can only take positive values $h>0$, we need to consider two different phases. The first phase, characterized by the magnetic field in the range $0\leq h \leq 1$, is called the symmetry broken phase (BP) \cite{PhysRevB.71.224420}. On the other hand, for $h>1$, the system is in the symmetric phase (SP). In both phases, the leading order terms contributing to the Hamiltonian in the thermodynamic limit can be obtained in terms of the bosonic creation and annihilation operators using the Holstein-Primakoff (HP) representation  of the spin components $J_{\alpha}$. This  Hamiltonian can now be diagonalized using a Bogoliubov transformation. For brevity, we leave out the standard details of this procedure \footnote{see \cite{PhysRevB.71.224420} for details of this diagonalization procedure.} and provide only the final form of the Hamiltonian in terms of the final set of creation and annihilation operators. 
 
In the BP, the final form of the Hamiltonian can be written in terms of the bosonic operators $\alpha_1, \alpha_1^{\dagger}$ as
\begin{align}
		\label{BP-Hamiltonian}
		\mathcal{H}_{BP}=2\sqrt{\left(1-h^{2}\right)\left(1-\gamma\right)}
		\left(\alpha_1^{\dagger}\alpha_1+\frac{1}{2}\right)~.
\end{align} 
However in the SP, we define the new set of bosonic operators $\alpha_2, \alpha_2^\dagger$ utilizing the Bogoliubov transformation and the final form of the Hamiltonian after the diagonalization is given by
\begin{align}
		\label{SP-Hamiltonian}
		\mathcal{H}_{SP}=2\sqrt{\left(h-1\right)\left(h-\gamma\right)}\left(\alpha_2^{\dagger}\alpha_2
		+\frac{1}{2}\right)~.
\end{align}
Thus, in both phases, the final form of the Hamiltonian is that of a simple harmonic oscillator (HO) written in terms of the corresponding creation and annihilation operators.
We also note that the Bogoliubov transformation is different in the two phases of the system. As the system approaches the QPT, the frequency of the corresponding oscillator from either phase goes to zero.

Introducing a position and its corresponding momentum operator we can write the above two Hamiltonians
as
\begin{equation}
	\label{oscillator}
	\mathcal{H}=\frac{1}{2}p_i^2+\frac{1}{2}\omega_{i}^2x_i^2~,~~~\text{where}~~~i=1,2~,
\end{equation}
 and 
\begin{equation}
	\label{coordinates}
	x_{i}=\frac{1}{\sqrt{2\omega_{i}}}\Big(\alpha_i^{\dagger}+\alpha_{i}\Big)~,
	~p_{i}=i\sqrt{\frac{\omega_{i}}{2}}\Big(\alpha_i^{\dagger}-\alpha_{i}\Big)~.
\end{equation}
In this section we  consider the quantum quenches  in this HO model which have frequencies same as the 
the LMG model in the thermodynamic limit. For $0\leq h \leq 1$ frequency of the 
oscillator is $\omega_{1}=2\sqrt{\left(1-h^{2}\right)\left(1-\gamma\right)}$, while for $h>1$ we have
$ \omega_{2}=2\sqrt{\left(h-1\right)\left(h-\gamma\right)}$. For convenience, we refer these two
cases as the BP and the SP respectively, and the point $h=1$ for which the frequency vanishes in both the
phases  is refereed as the critical point.

Before delving into the computation of the time evolved state after the quench, and subsequently to the calculation of the complexity, we emphasize that, in this section we are not considering the quench in the LMG model in \cref{LMGHamiltonian}. We rather consider quench in a HO model whose frequency in the cases $h<1$ and $h>1$ coincides respectively with that of the LMG model in the BP and the SP at the thermodynamic limit. The time evolution of the SC after a quantum quench in the LMG model is discussed in \cref{SC_LMG_Exact}. There we observe that the SC of the oscillator model presented here shows all the characteristics of the exact result. Here the quench of the exact LMG model is not considered in this section since in the BP of the LMG model, it is relatively difficult to calculate the time evolved state after a general quench where the magnetic field has different values before and after the quench. In case of the HO model we consider in this section, the time evolved state can be calculated using Lie algebraic methods and hence the time evolution of both SC and spread entropy (considered in the next section) are analytically tractable.

\subsection{Time evolved state after a sudden quantum quench}\label{quench}

The quench protocol we consider is the following. The HO in \cref{oscillator} is assumed to be in the ground state with the values of the parameter $h_i$ and $\gamma_i$. The system can either be in the BP or the SP, depending on the initial magnetic field $h_i$. At $t=0$, we suddenly change the magnetic field to a new value $h_f$ so that the new parameter $h_f$ still characterizes the same phase as the initial one. Thus the phase before and after the quench is assumed to remain the same. We can also suddenly change the anisotropy parameter $\gamma_i$ to a new value $\gamma_f$. However, in the numerical calculations below, we mostly keep it to be a constant.

To find out the time evolved state after the quench (eq. (\ref{psi_ft})), we need to write
down the Hamiltonian $\mathcal{H}_f$ after the quench in terms of the operators
$\alpha_{ji}$ and $\alpha^\dagger_{ji}$ ($j=1,2)$ before the quench. This can be
accomplished by realizing that the bosonic operators before and after the quench
are related by a Bogoliubov transformation \cite{Ali:2018fcz}
\begin{equation}\label{Bogoliubov}
	\left(
	\begin{array}{ccc}
		 \alpha_{jf} \\
		\alpha^\dagger_{jf} \\ 
	\end{array}
	\right)=\left(
	\begin{array}{ccc}
		\mathcal{U}_j &\mathcal{V}_j  \\
		\mathcal{V}_j  & \mathcal{U}_j \\ 
	\end{array}
	\right) \left(
	\begin{array}{ccc}
		\alpha_{ji} \\
		\alpha^\dagger_{ji} \\ 
	\end{array}
	\right)~,
\end{equation}
where the Bogoliubov coefficients are $~\mathcal{U}_j=\frac{\omega_{jf}+\omega_{ji}}{2\sqrt{\omega_{ji}\omega_{jf}}}~,~
\mathcal{V}_j=\frac{\omega_{jf}-\omega_{ji}}{2\sqrt{\omega_{ji}\omega_{jf}}}$.
Here the subscript $j$ indicates which of the two cases ($h>1$ or $0\leq h \leq 1$)  we are considering. In terms
of the operators before the quench, the post-quench Hamiltonian can then be written as
\begin{equation}
	\mathcal{H}_f=2\omega_{jf}\Big[\mathcal{U}_j\mathcal{V}_jK^{+}+\big(\mathcal{U}_j^2+
	\mathcal{V}_j^2\big)K^0+\mathcal{U}_j\mathcal{V}_jK^{-}\Big]~.
\end{equation}
Here the operators $K^{+},K^{0}$ and $K^{-}$ are the generators of the $su(1,1)$ Lie algebra, and are
related to creation and annihilation operators before the quench through the  following relations
\begin{equation}\label{su11op}
	K^{+}=\frac{1}{2}\alpha^\dagger_{ji}\alpha^\dagger_{ji}~,~ 
	K^{0}=\frac{1}{4} \Big(\alpha^\dagger_{ji}\alpha_{ji} + \alpha_{ji} \alpha^\dagger_{ji}  \Big)~,
	~\text{and}~ K^{-}=\frac{1}{2}\alpha_{ji}\alpha_{ji}~.
\end{equation}
The generators $K_i$  provide a single-mode bosonic representation of the $su(1,1)$ Lie algebra,
 and satisfy the usual commutation relations
 \begin{equation}
 	\big[K_{+},K_{-}\big]=-2K_{0}~,~~\big[K_{0},K_{\pm}\big]=\pm K_{\pm}~.
 \end{equation}
The corresponding Casimir operator, defined as 
\begin{equation}
	K^2=K_0^2-\frac{1}{2}\big(K_+K_-+K_-K_+\big)~,
\end{equation} 
 commutes with all the three generators of the algebra, and satisfies the following eigen value equation
\begin{equation}\label{Casimir}
	K^2\ket{m,k}=k(k-1)\ket{m,k}~.
\end{equation}
Here the constant $k$ is the Bargmann index of the algebra, and $m$ takes values $0,1,2\cdots$. For the single-mode bosonic representation of  $su(1,1)$ Lie algebra given above, the Bargmann index $k$ can take values $1/4$ or $3/4$ (see \cite{Gerry:91}). For $k=1/4$, the basis corresponding to a unitary irreducible representation of the algebra is the set of states with an even number of bosons. In this paper, we always consider $k$ to be $1/4$. The operations of the generators $K_i$ on the states $\ket{m,k}$ are given by the usual formulae, which can be found, for example, in \cite{Gerry:91}.

Now using the well known decomposition relations for the $SU(1,1)$ group elements, the time evolved state after the quench can be written as \footnote{See \cite{Ban:93} for a derivation of this formula and some other such decompositions in more general scenarios.} 
\begin{equation}\label{decomp_LMG}
	\begin{split}
	\ket{\Psi_f(t)}=e^{-i\mathcal{H}_f t} \ket{\psi(t=0)}=
	\exp \bigg[- 2 it  \omega_{jf}\Big(\mathcal{U}_j\mathcal{V}_jK^{+}+\big(\mathcal{U}_j^2+
	\mathcal{V}_j^2\big)K^0+\mathcal{U}_j\mathcal{V}_jK^{-}\Big)\bigg]\ket{0}\\
	=\exp \Big[\mathbf{X_+}K^{+}\Big] \exp \Big[\ln \mathbf{X_0}~K^{0}\Big]
	\exp \Big[\mathbf{X_-}K^{-}\Big]\ket{0}~.
\end{split}
\end{equation}
Here $\ket{0}$ indicates the ground state of the HO ($\alpha_{ji} \ket{0}=0$) and therefore, also annihilated by the operator $K^{-}$ (see the definition in eq. \eqref{su11op}). We assume that this is the state
	the system was  before the quench.
The three  functions $\mathbf{X}_i$s are given by
\begin{equation}
	\mathbf{X}_{\pm}=\Big(\frac{\mathbf{x}_{\pm}}{\Theta}\Big) \sqrt{\mathbf{X}_0}\sinh \Theta~,~
	\mathbf{X}_0=\Big(\cosh \Theta - \frac{ \mathbf{x}_0}{2 \Theta} \sinh \Theta\Big)^{-2}~,
\end{equation}
with 
\begin{equation}
	\mathbf{x}_{\pm}=-2it\omega_{jf}\mathcal{U}_j\mathcal{V}_j~,~~
	\mathbf{x}_0=-2it\omega_{jf}\big(\mathcal{U}_j^2+\mathcal{V}_j^2\big)~,~~
	\text{and}~~\Theta^2=\frac{1}{4}\mathbf{x}_0^2-\mathbf{x}_{+}\mathbf{x}_{-}~.
\end{equation}
It can  be seen that since the Hamiltonian is written in terms of the creation and annihilation operators before the quench, which are the elements of the $su(1,1)$ algebra, the time evolved state is a general $su(1,1)$ coherent state. For the case of $k=1/4$, we are considering the scenario when $\mathbf{x}_0=0$ and the state reduces to the usual squeezed vacuum state.

Using the decomposed time evolved state in eq. (\ref{decomp_LMG}), we can obtain the  
auto-correlation function to be
\begin{equation}
	\mathcal{S}(t)=\braket{\Psi_f(t) | \psi(t=0)}=\Big[\cosh \bar{\Theta} + 
	\frac{ \mathbf{x}_0}{2 \bar{\Theta}} \sinh \bar{\Theta}\Big]^{-1/2}~,
	~\text{with}~	\bar{\Theta}^2=\frac{1}{4}\bar{\mathbf{x}}_0^2-\bar{\mathbf{x}}_{+}\bar{\mathbf{x}}_{-}~,
\end{equation}
and the $\bar{\mathbf{x}}_j$s are the complex conjugates of $\mathbf{x}_j$s. Using the expressions for the
$\mathbf{x}_j$s given above, we can rewrite the expression for the auto-correlation as 
\begin{equation}\label{St}
	\mathcal{S}(t)=\Big[\cos  \big(\omega_{jf} t\big) 
	-i \big(\mathcal{U}_j^2+\mathcal{V}_j^2\big) \sin \big(\omega_{jf} t\big)\Big]^{-1/2}~.
\end{equation}
This auto correlation function is utilized to obtain Lanczos coefficients in the following subsection.

\subsection{The Lanczos coefficients and the spread complexity}\label{Lanczos}
In this subsection, we first compute the Lanczos coefficients using the algorithm described in sec. \ref{procedure} using the auto-correlation function given in eq. (\ref{St}). These coefficients can compactly be written as \footnote{Apart from quantities such as, $\mathcal{U}_j, \mathcal{V}_j$ and $\omega_{jf}$, whose expressions are different in the two phases, we mostly omit the index $j$ in other quantities (such as the Lanczos coefficients) from now on.}
\begin{eqnarray}
	\begin{split}
	a_n=\Big(2n+\frac{1}{2}\Big)\big(\mathcal{U}_j^2+\mathcal{V}_j^2\big)\omega_{jf}~,~
	\text{with} ~~n=0,1,2,3 \cdots~, 	\\~\text{and}~~~~
	b_l=\sqrt{2(2l^2-l)}\mathcal{U}_j|\mathcal{V}_j| \omega_{jf}~,~\text{with}~~l=1,2,3 \cdots ~.
\end{split}\label{lanczos}
\end{eqnarray}
Notice that the modulus of the quantity $\mathcal{V}_j$ appears in the expression for $b_l$. This is because for the particular quench model we are considering, $\mathcal {V}_j$ is actually negative in both phases. We explicitly assume that the initial state is always away from the criticality, while the state after the quench is close to the critical point.
However, for the case of critical quench considered later, the state after the quench is at the critical point. Since the frequency goes to zero at the critical point, it can be seen from (\ref{Bogoliubov}) that $\mathcal{V}_j$ is negative for such quenches. In this context, we can consider some particular quench protocols where the initial state is close to the QPT. Here also, we can study the evolution of the SC when the initial state is gradually moved towards the critical point. In that case, $\mathcal{V}_j$ changes its sign when $\omega_{ji}$ crosses $\omega_{jf}$ through successive quenches.

Now using  eqs. (\ref{Krylov-basis}) and (\ref{lanczos}), we can find out the elements of the Krylov basis. It can be readily checked that these are the eigen states $\ket{m,k}$ of the Casimir operator $K^2$ defined in eq. (\ref{Casimir}). For our case, as emphasized before, the state of the system before the quench is the first state of the Krylov basis which makes the subsequent procedure  easier to track analytically.

Next we proceed to the calculation of $\phi_{n}(t)$s and subsequently, the  SC. Following the procedure described in sec. \ref{procedure}, we obtain these to be given by
\begin{equation}\label{phi_n}
	\phi_{n}(t)=\mathcal{N}_n \phi_0 (t)\frac{\mathcal{G}_1(t)^n}{\mathcal{G}_2(t)^n}~
	=\mathcal{N}_n \phi_0 (t)\mathcal{G}(t)^n~,
\end{equation}
where $\phi_0=\mathcal{S}(t)^{*}$, and the time-dependent function $\mathcal{G}(t)$ is defined as 
\begin{equation}
	\begin{split}
		\mathcal{G}(t)=\frac{\Big(\omega_{jf}^2-\omega_{ji}^2 \Big)\sin (\omega_{jf} t)}
		{\Big(\omega_{jf}^2+\omega_{ji}^2 \Big)\sin (\omega_{jf} t)-2i \omega_{jf}\omega_{ji} \cos (t\omega_{jf})}~.
	\end{split}
\end{equation}
The quantities $\mathcal{N}_n$s appearing above are numerical constants. Here we record the first few values  for convenience: $\mathcal{N}_0=1, ~\mathcal{N}_1=\frac{1}{\sqrt{2}}, ~\mathcal{N}_2=\frac{\sqrt{3}}{2\sqrt{2}},
~ \mathcal{N}_3=\sqrt{\frac{5}{16}},~\mathcal{N}_4=\frac{\sqrt{35}}{8\sqrt{2}}~,\cdots$

The first important point we note from the expressions for the functions $\phi_{n}(t)$s, which  hugely simplify 
the calculation of the complexity is that,  the ratio of two successive $\phi_{n}(t)$s are related by the following relation
\footnote{With the help of eq. (\ref{ratio_phin}) it can be easily verified that the condition $\sum_{n}|\phi_n(t)|^2=1$ is satisfied as well. }
 \begin{equation}\label{ratio_phin}
 	\frac{\phi_{n+1}(t)}{\phi_{n}(t)}=\frac{\mathcal{N}_{n+1}}{\mathcal{N}_n}\mathcal{G}(t)~=
 	~\sqrt{\frac{2n+1}{2n+2}}\mathcal{G}(t)~.
 \end{equation}
With this observation, it is  easy to write down an exact analytical expression for the SC given in eq. (\ref{spread_complexity}) as 
\begin{equation}\label{SCsum}
	\mathcal{C}(t)=2|\phi_1(t)|^2\sum_{n=0}^{\infty} \frac{(n+1) (2n+1)!!}{(2n+2)!!}\mathcal{F}^{n}(t)
	   =\frac{|\phi_1(t)|^2 }{\big(1-\mathcal{F}(t)\big)^{3/2}}~,\text{where}~~
	   \mathcal{F}(t)=|\mathcal{G}(t)|^2~.
\end{equation}
Here the expression for the time-dependent function $\mathcal{F}(t)$ is given by
\begin{equation}
	\mathcal{F}(t)=\frac{\big(\omega_{jf}^2-\omega_{ji}^2 \big)^2\sin ^2(\omega_{jf} t)}
	{\big(\omega_{jf}^2+\omega_{ji}^2 \big)^2\sin^2(\omega_{jf} t)+4\omega_{jf}^2\omega_{ji}^2 \cos^2(t\omega_{jf})}~.
\end{equation}

From the expression for the complexity, we see that for it to be well defined, the function $\mathcal{F}(t)$, and hence the ratio of the modulus squared values of two successive $\phi_n(t)$s must be smaller than unity
at all times. This also provides the reason for the convergence of the infinite sum appearing in the expression for the complexity. Furthermore, the time dependence of the complexities after the quench in both phases of the oscillator are essentially determined by the ratio of the modulus squared of successive $\phi_n$s and the modulus of the coefficient of $\ket{K_{1}}$ in the expansion of the time evolved state.
 
Finally, using the expression for $\phi_{1}$ from eq. (\ref{phi_n}), the formula for the SC reduces to
\begin{equation}\label{complexity_2}
	\mathcal{C}_j(t)=\frac{\big(\omega_{ji}^2-\omega_{jf}^2\big)^2}{8 \omega_{jf}^2 \omega_{ji}^2}
	\sin ^2 \big(\omega_{jf} t\big)~.
\end{equation}

This formula is valid in both phases of the ground state (with $j=1,2$ for the broken and the SP respectively). However, the exact dependence of the SC on the parameters of the system ($h, \gamma$), which are changed during the quench, are different
in two phases. With the expressions for the frequencies $\omega_{jf}$ and $\omega_{ji}$ given 
in eqs. (\ref{BP-Hamiltonian}) and (\ref{SP-Hamiltonian}) we now separately study the time evolution of the SC after the quantum quench in the two phases.

\subsubsection*{Evolution of  the spread complexity for $0\leq h \leq 1$}
We first assume that the system is prepared at time $t=0$ in the ground state with a magnetic field 
in the range $0\leq h \leq 1$.  The value of $h_i$ is chosen in such a way that it is away from the critical value of the magnetic field, i.e. $h_i<h_c$. At $t=0$, the parameters $h_i, \gamma_i$ are suddenly changed to new values $h_f, \gamma_f$, and subsequently, the evolution of the system is governed by the new Hamiltonian. We take the magnetic field after the quench $h_f$ to be greater than the initial value $h_i$, but smaller than the critical value $h_c$, i.e. $h_i<h_f<h_c$. Thus, the state after the quench is still in the ground state at the BP, but is closer to  criticality as compared to the initial state. The special case of critical quench characterized by  $h_f = 1$ will be discussed separately in subsection  \ref{critical_quench}.

\begin{figure}[h!]
	\centering
	\includegraphics[width=0.65\textwidth]{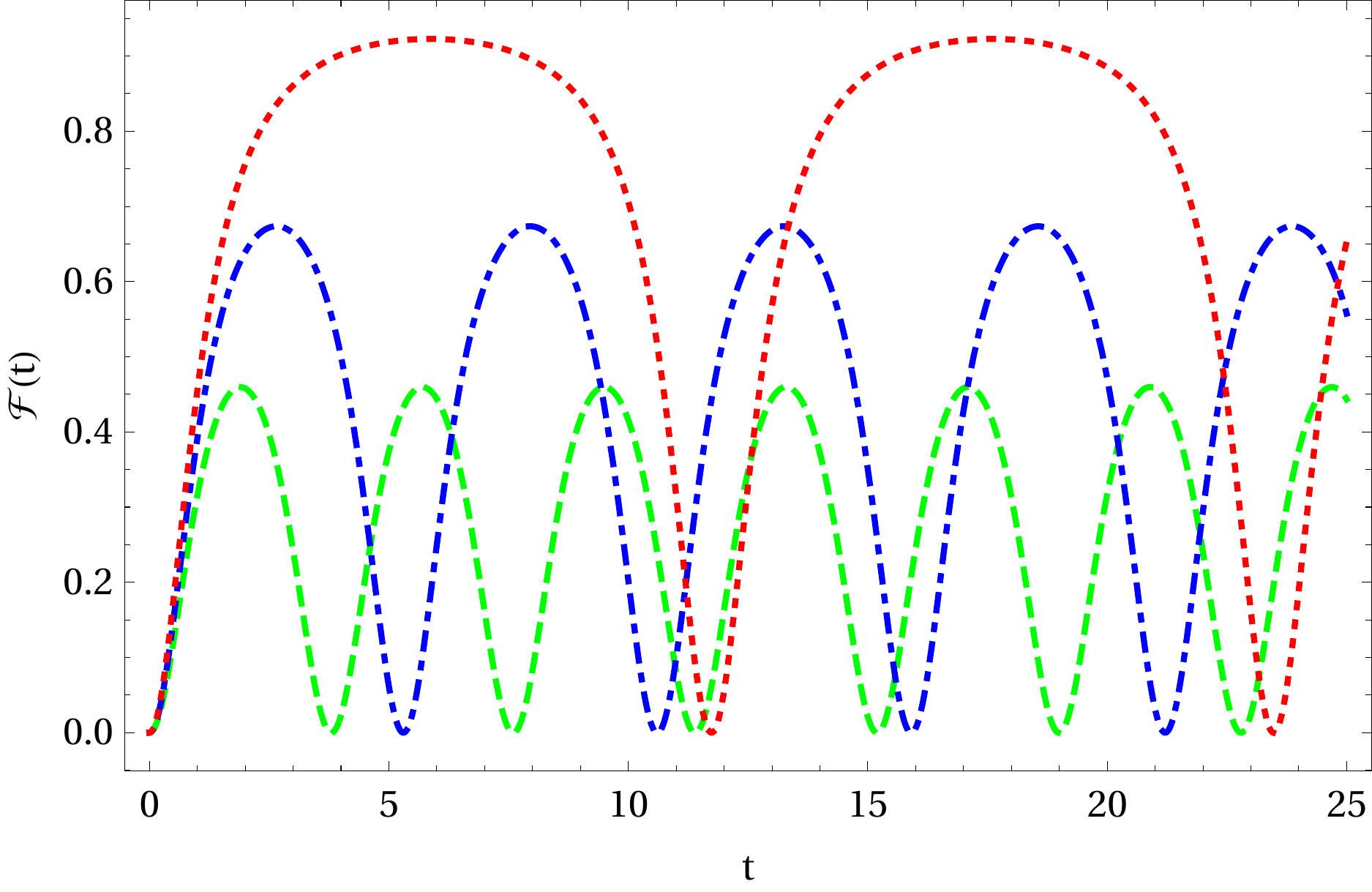}
	\caption{Variation of $\mathcal{F}(t)$ with time in the BP for different post-quench magnetic fields. Here $h_f=0.9$ (green), $h_f=0.95$ (blue) and $h_f=0.99$ (red) respectively. The parameter $\gamma$ has a fixed value $0.1$, and $h_i=0.5$.}
	\label{fig:F_broken}
 \end{figure}

We first study the behavior of the function $\mathcal{F}(t)$ when $h_f$ is gradually taken closer to the critical point. Fig.~\ref{fig:F_broken} exhibits oscillatory behavior of the function $\mathcal{F}(t)$ for different values of the post-quench magnetic field $h_f$  in the BP, by keeping $\gamma$ fixed. Here we consider the magnetic field after the quench to be  close to the critical value $h_c=1$, but always less than one. One can see that when $h_f$ gradually approaches the critical point, the amplitude and the time period of the oscillation increase. However, as expected, the function $\mathcal{F}(t)$ is always lower than one.

    \begin{figure}[h!]
  	\centering
  	\includegraphics[width=0.65\textwidth]{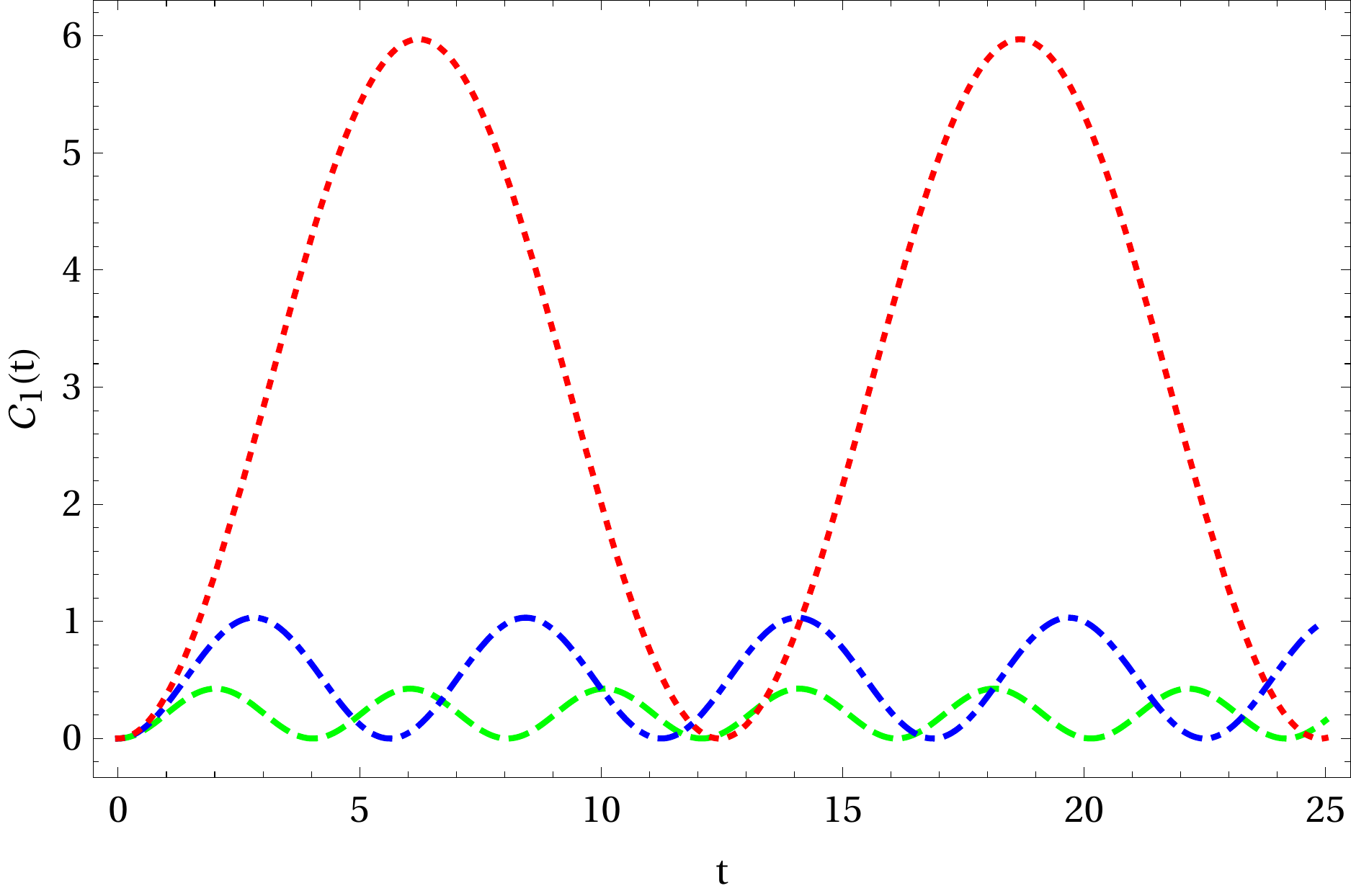}
  	\caption{Evolution of $\mathcal{C}_1(t)$ with time  in the BP for different post-quench magnetic fields. The parameter values and color coding are the same as those of  Fig. \ref{fig:F_broken}. }
  	\label{fig:C_broken}
  \end{figure}

The expression for the SC after the quench in the BP is given by 
\begin{equation}\label{complexity_broken}
	\mathcal{C}_1(t)=\frac{\big[(1-h_i^2)(1-\gamma_i)-(1-h_f^2)(1-\gamma_f)\big]^2}
	{8 (1-h_i^2)(1-\gamma_i) (1-h_f^2)(1-\gamma_f)}
	\sin ^2 \Big(2 \sqrt{(1-h_f^2)(1-\gamma_f)} t\Big)~.
\end{equation}
This formula is written down for the most general quench scenario 
with $h_f\neq h_i$ and $\gamma_i\neq \gamma_f$. In the special
case when $\gamma_i = \gamma_f$ it can be seen that $\gamma$ only affects the time period of the oscillations, and not the magnitude. Below we mostly focus on this special case, since in this case the SC  carries all the typical characteristics of the general quench.

Time evolution of the SC for the quench protocols considered in Fig.~\ref{fig:F_broken} is
shown in Fig.~\ref{fig:C_broken}.\footnote{Unless stated otherwise, we henceforth always set $\gamma_f=\gamma_i=0.1$ in the plots and any reference of other numerical quantities.}
The SC oscillates with time, and the oscillation amplitude and time period gradually increase as the post-quench magnetic field is taken closer to the criticality. 
 
From the formula (\ref{spread_complexity}), we see that the SC is a weighted sum of the squared modulus of the $\phi_{n}$s. Now, one can notice that the SC increases as the post-quench magnetic field is taken closer to the critical point. Hence, even though we have the exact expression for the infinite sum in eq. (\ref{spread_complexity}), it is interesting to quantify how many  $\phi_{n}$s ``significantly''
\footnote{The exact mathematical meaning  of ``significantly'' contributing elements of the Krylov basis is defined shortly. }
contribute to the total complexity as the magnetic field moves closer to the critical point. 

\begin{figure}[h!]
	\centering
	\includegraphics[width=0.65\textwidth]{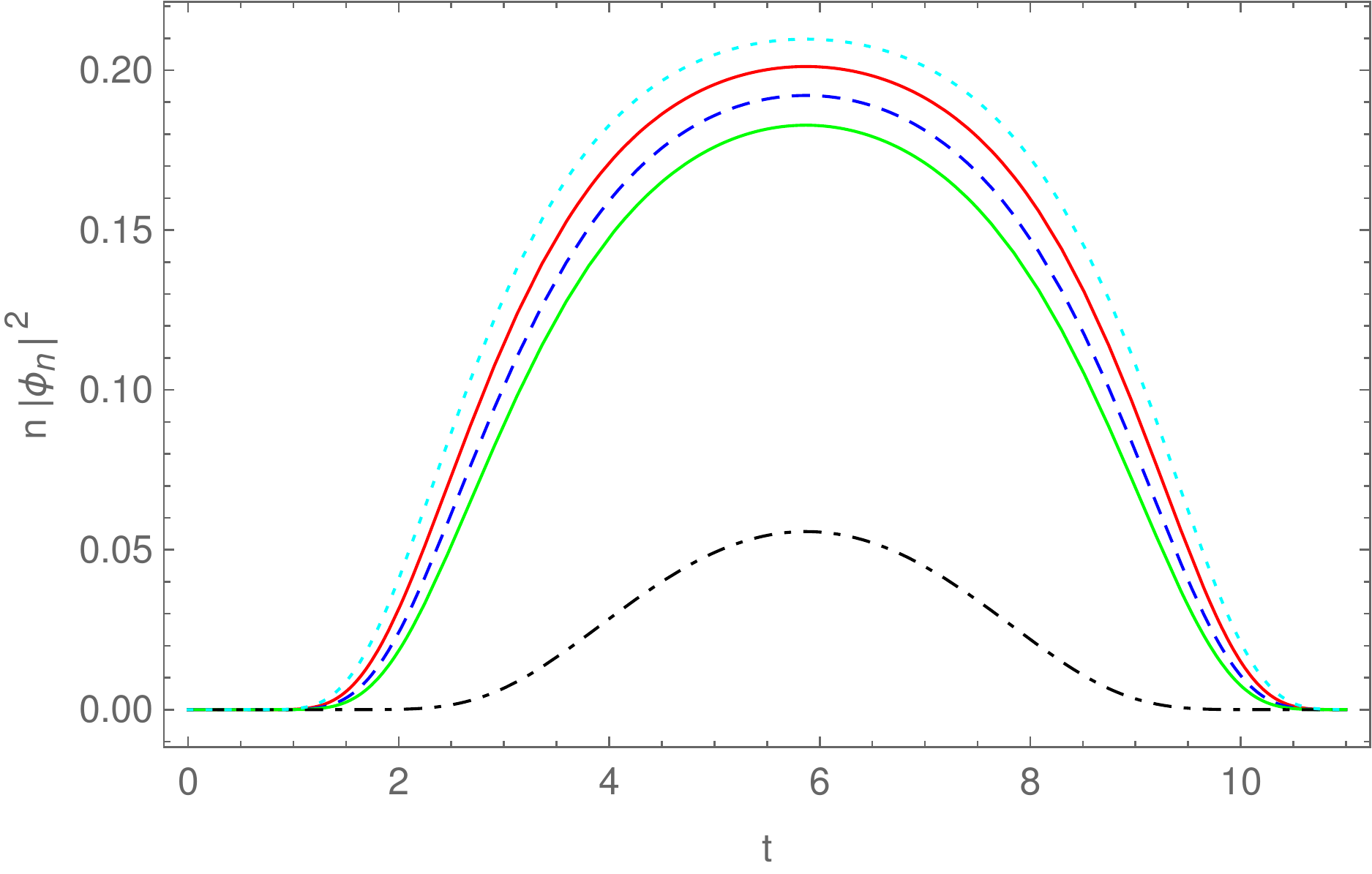}
	\caption{Plot of the individual contributions to the SC sum in eq. \eqref{SCsum}. Here $n=9$ (cyan),
	 $n=10$ (red), $n=11$ (blue), $n=12$ (green), $n=30$ (black). The parameter values are $h_i=0.1, h_f=0.99, 
	 \gamma_i=\gamma_f=0.1$. For these values of the parameters, the 
	 individual contributions towards the total sum start to decrease from $n=10$. }
	\label{fig:terms}
\end{figure}

In the BP of the system, when the final magnetic field is far from the critical point, the SC $\mathcal{C}_1(t)$ converges rapidly towards its exact value. Hence, only the first few values of $n$ contribute to the complexity. For example, when $h_f=0.5$, and $h_i=0.1$ only the terms $n=1$ and $n=2$ have  contribution greater than the value $0.00025$ towards $\mathcal{C}_1(t)$. On the other hand, when  $h_f$ is closer to $1$, there is a crossover in the magnitude of individual contributions in eq. (\ref{spread_complexity}), i.e., there exists a maximum value of $n$ such that the quantities $(n+j)|\phi_{(n+j)}(t)|^2$ are smaller than  $n|\phi_{n}(t)|^2$ for all $j\geq1$. Since the quantities $n|\phi_{n}(t)|^2$ are functions of time, by the statement `$(n+j)|\phi_{(n+j)}(t)|^2$ are smaller than  $n|\phi_{n}(t)|^2$', we mean that the maximum values of $(n+j)|\phi_{(n+j)}(t)|^2$ are smaller than the maximum value of $n|\phi_{n}(t)|^2$. There may be other values of time where $(n+j)|\phi_{(n+j)}(t)|^2$ are actually greater than $n|\phi_{n}(t)|^2$. However, for sufficiently higher values of $j$, this statement is applicable for all times.
As an  example, when we set $h_f=0.99$, with $h_i=0.1$, we notice that the individual contribution to the complexity starts to decrease from $n=10$.\footnote{Though the individual terms start to decrease, they may still contribute to the
total SC. This is best characterized by the quantity, effective number of elements of the Krylov basis, 
introduced  below.} This phenomenon can be clearly seen from plots in Fig. \ref{fig:terms}, where we have shown 
that the contribution of the $n=11$th term is smaller than $n=10$th term (for the values of the parameters 
mentioned above and the caption of this figure), and all the higher $n$ terms have gradually smaller contributions.

Now we want obtain the ``effective number of elements'' $n=N_{eff}$ of the Krylov basis,
those contribute to the sum in $\mathcal{C}_1(t)$ for different values of $h_f$ close to criticality, up to a
predefined cut off. We quantify $N_{eff}$ such that the inequality  in eq. \eqref{N_eff}
is satisfied by $\mathcal{C}_1(t)$ in the BP.

\begin{figure}[h!]
	\centering
	\includegraphics[width=0.7\textwidth]{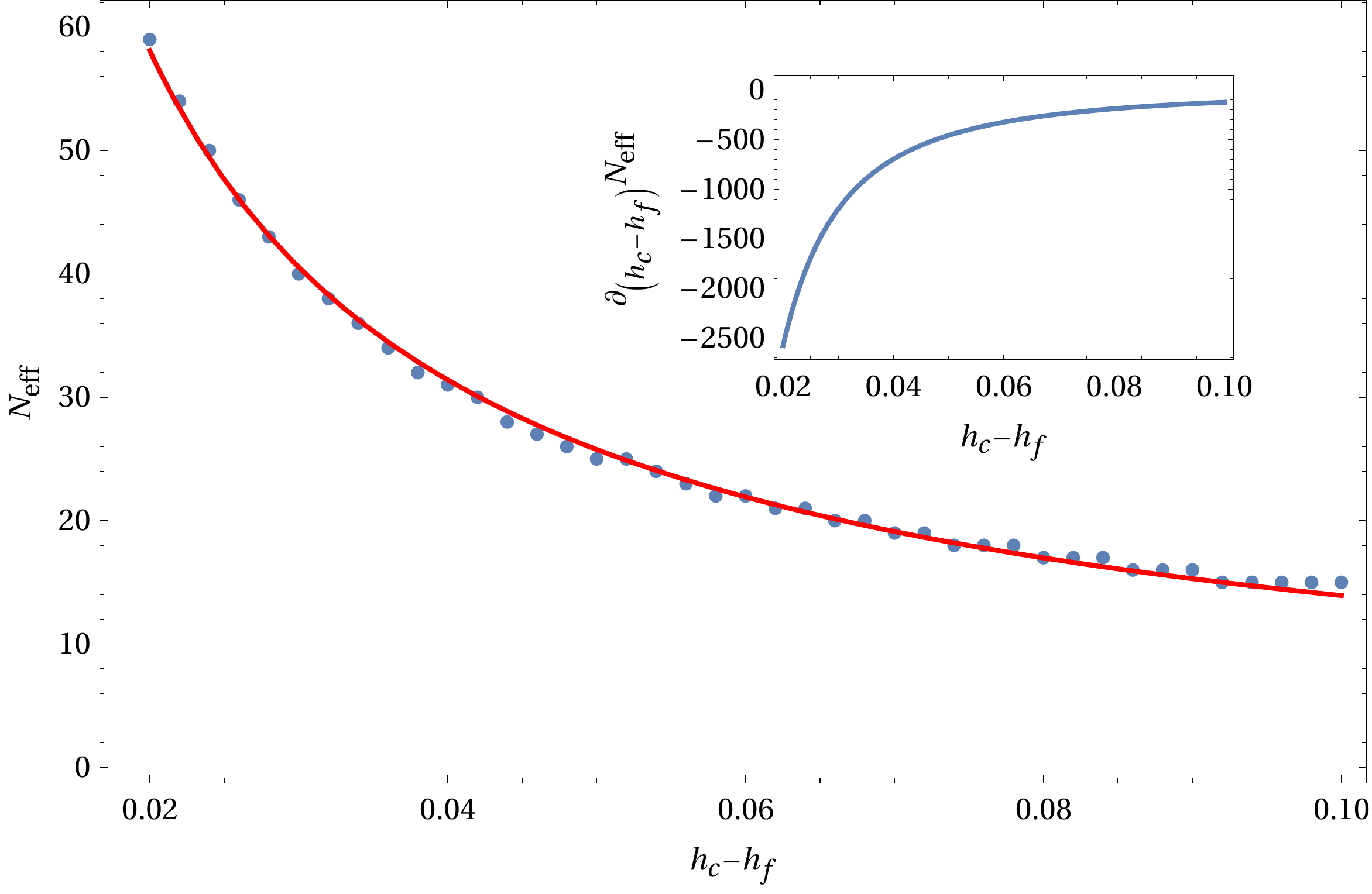}
	\caption{Plot of $N_{eff}$ with $h_f$ close to the criticality in the BP. The blue dots are numerical data which is fitted with the red curve where the equation of the red curve is given in eq. (\ref{fit}). The derivative of $N_{eff}$ with respect to $h_f$ is shown in the inset. We set $\epsilon=0.001$ and $h_i=0.1$.}
	\label{fig:eff_broken}
\end{figure}

In Fig.~\ref{fig:eff_broken} we plot $N_{eff}$,\footnote{As we have mentioned in the discussion after eq. \eqref{N_eff},
	since $\mathcal{C}(t)$ is a function of time, the criterion in eq. (\ref{N_eff}) need to be applied at some 
	specified  instant of time. Here, we have imposed it at the instant when the complexity is maximum.}
  which satisfy the criterion in eq. (\ref{N_eff}),
as a function of $\Delta h=h_c-h_f$, when $h_f$ is taken close to the criticality. As can be anticipated from the discussion above, $N_{eff}$ increases with $h_f$ approaching the criticality. We can fit the data of $N_{eff}$ for different values of $h_f$ with reasonable accuracy with the following fitting function 
\begin{equation}\label{fit}
	N_{eff}(h_f)=\frac{n_1}{|h_c-h_f|^{n_2}}~,
\end{equation}
where in the BP, the numerical coefficients have values $n_1 \approx 1.808$ and $n_2 \approx 0.887$. This fitting function is shown in Fig. \ref{fig:eff_broken} with the red curve, 
and  the exact result obtained numerically are shown with blue dots. 
Furthermore, to quantify the growth of $N_{eff}$ with $h_f$, we also show the plot of the 
derivative of $N_{eff}(h_f)$ in the inset, which approaches zero sharply towards the criticality.

\begin{figure}[h!]
		\centering
	\includegraphics[width=0.65\textwidth]{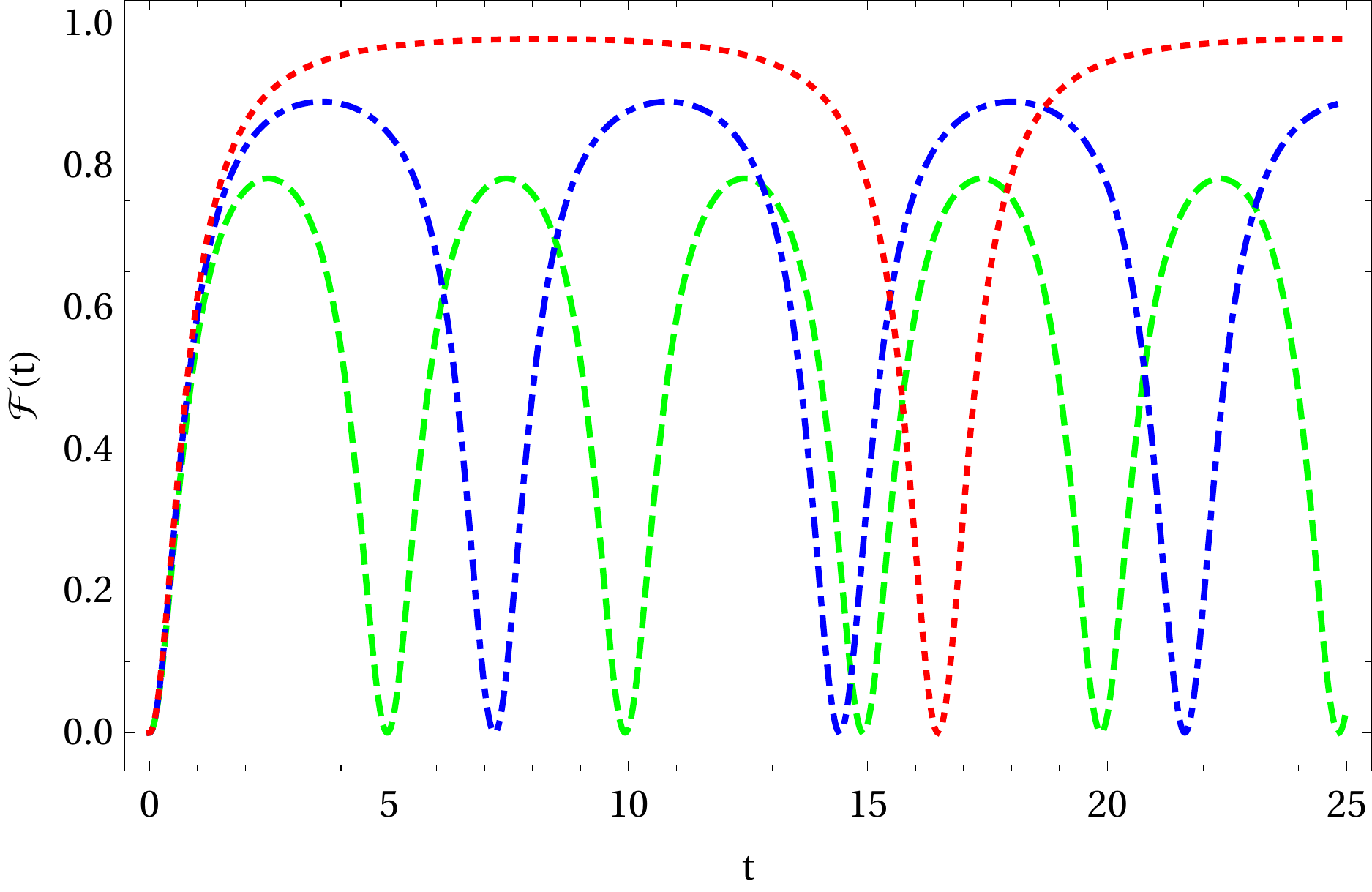}
	\caption{Variation of $\mathcal{F}(t)$ with time in the SP of the system 
		for different post-quench magnetic fields.
		Here $h_f=1.1$ (green), $h_f=1.05$ (blue) and $h_f=1.01$ (red) respectively. The parameter
		$\gamma$ has a fixed value $0.1$, and $h_i=1.5$. }
	\label{fig:F_symmetric}
\end{figure}

In this context we note that,  if we take $\epsilon$ smaller than the value $0.001$ we have used above,
there will be  a saturation in the estimation of $N_{eff}$ close to the criticality, i.e. there will be  ranges of $h_f$ for each of which the quantity $N_{eff}$ will be the same.

\subsubsection*{Evolution of the  spread complexity for $h>1$}

Now we assume that the system is prepared in the ground state of the oscillator with magnetic field $h_i>1$. The magnetic field after the quench is smaller than $h_i$ and larger than $h_c$, i.e. $1< h_f<h_i$. Thus the state of the system after the quench is still in the ground state  of the SP but closer to  criticality.

Fig.~\ref{fig:F_symmetric} exhibits the time variation of $\mathcal{F}(t)$ for different values of the post-quench magnetic field $h_f$, by keeping $\gamma$ fixed in SP. The qualitative behavior of this function is similar to that of the one in the BP. When the magnetic filed values approach  the critical point, the  peak value of $\mathcal{F}(t)$ gets closer to $1$, but always remains smaller than unity.

\begin{figure}[h!]
		\centering
		\includegraphics[width=0.65\textwidth]{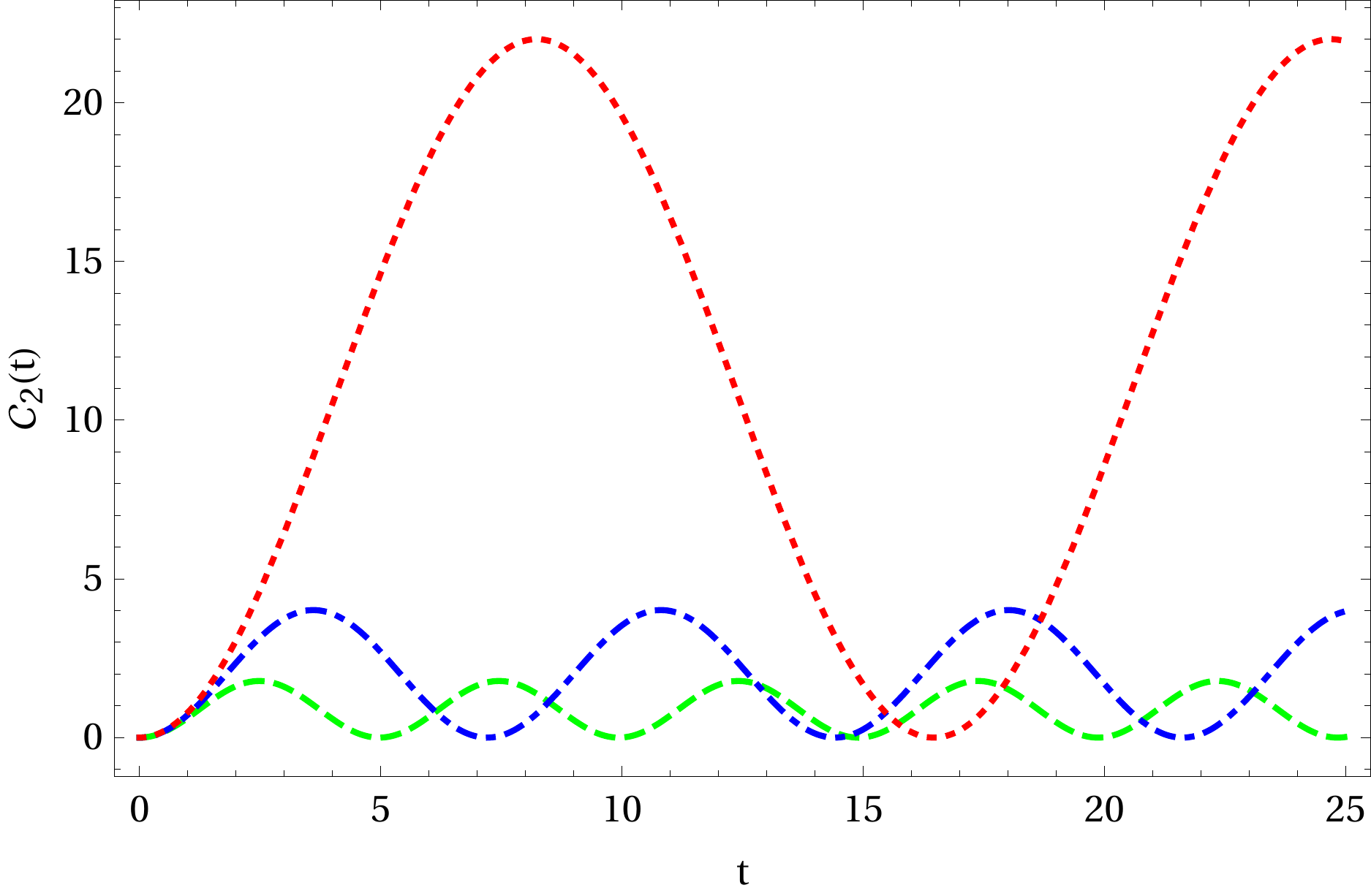}
		\caption{Evolution of $\mathcal{C}_2(t)$ with time for different post-quench magnetic fields in the SP.	The parameter values and color coding  are the same as in Fig. \ref{fig:F_symmetric}. }
		\label{fig:C_symmetric}
\end{figure}

For quantum quenches in the SP, the expression for the SC can be written as
\begin{equation}\label{complexity_symmetric}
	\mathcal{C}_2(t)=\frac{\big[(h_i-1)(h_i-\gamma_i)-(h_f-1)(h_f-\gamma_f)\big]^2}
	{8 (h_i-1)(h_i-\gamma_i) (h_f-1)(h_f-\gamma_f)}
	\sin ^2 \Big(2 \sqrt{(h_f-1)(h_f-\gamma_f)} t\Big)~.
\end{equation}
The dependence of $\mathcal{C}_2(t)$ on $\gamma$ in this case is different from $\mathcal{C}_1(t)$, namely, for the 
SP even when  $\gamma_f=\gamma_i$ the anisotropy parameter can affect the magnitude of the 
SC.

Time evolution of $\mathcal{C}_2(t)$ in the SP for quenches with different final values of the magnetic field near criticality, is shown in Fig.~\ref{fig:C_symmetric}. As $h_f$ approaches towards the critical point, both the amplitude and the time period of oscillations gradually increase.  

\begin{figure}[h!]
	\centering
	\includegraphics[width=0.7\textwidth]{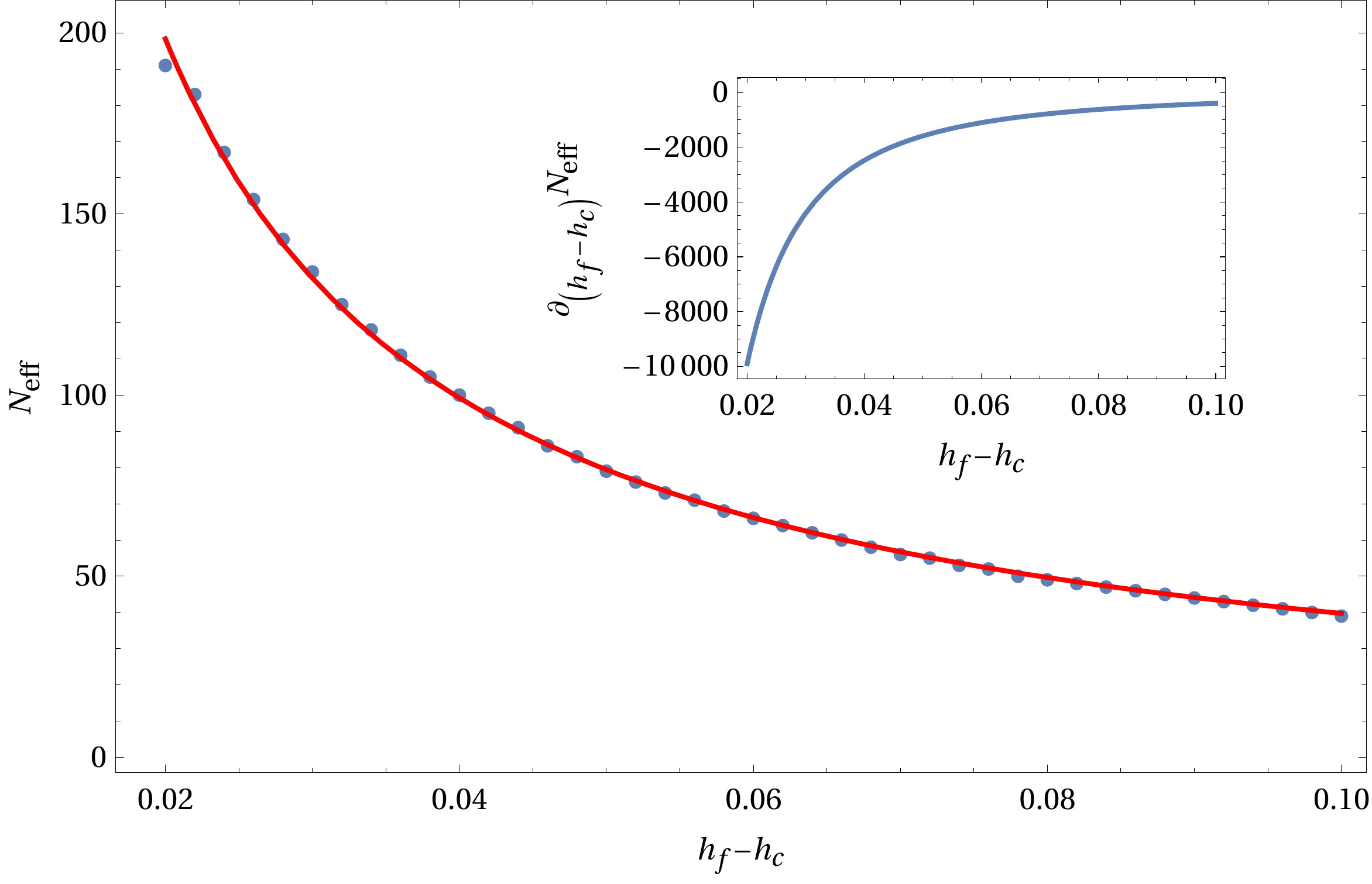}
	\caption{Plot of $N_{eff}$ with $h_f$ close to the criticality in the SP. The fit of the numerical data (blue dots) with eq.~(\ref{fit}) (red curve). The derivative of $N_{eff}$ with respect to $h_f$ is shown in the inset. Here we set $\epsilon=0.001$ and $h_i=1.9$.}
	\label{fig:eff_symmetric}
\end{figure}

Similar to the BP, we can quantify the number of terms contributing to the SP expansion
up to a preassigned cutoff value. One can check that for magnetic fields which are equally distant from the critical value $h_c=1$ in the two phases of the ground state, the number $N_{eff}$ that satisfy an equation  similar to eq. (\ref{N_eff}) with $\mathcal{C}_1(t)$ replaced by $\mathcal{C}_2(t)$, is greater in the SP than in the BP. Similarly, the crossover of the individual contributions towards the 
complexity, as  described above for the BP, occurs in the SP as well, however for comparatively larger values of $n$. 

The variation of $N_{eff}$ with respect to $\Delta h=h_f-h_c$,  taken gradually closer to the critical point, is shown in Fig.~\ref{fig:eff_symmetric}, along with the fitting function in eq.~(\ref{fit}) indicated by the red dashed curve. Here the numerical constant have values $n_1 \approx 3.98$ and $n_2 \approx 0.999$ for the best fit of the numerical data. From the derivative of $N_{eff}$ shown in the inset of this figure we see that, compared to the BP, close to the criticality,  $N_{eff}$ more sharply goes to zero. Thus we conclude that, as we approach the critical point using quenches in the SP, the time evolved wavefunctions after quench spread over a larger number of elements of the Krylov basis compared to the symmetrically performed quenches in the BP.

Before concluding this subsection we note that, similar to the behavior of the quantity $N_{eff}$
considered in this paper, it is known in the literature that different  information
theoretic quantities those have been used to probe QPT in the LMG model have different scaling exponent
around the critical point when considered in the two phases of the system. For example, behavior of the 
 the quantity fidelity susceptibility around the critical point in the LMG has been considered in \cite{PhysRevE.78.032103},
 and it was shown that  the intensive fidelity susceptibility scales around the critical point as
 \begin{equation}
 	\chi_F \propto \frac{1}{|h_f-h_c|^\nu}~,
 \end{equation}
where the critical exponent $\nu=1/2$ when $h_f$ is in the BP of the system, while $\nu=2$ when $h_f$ is in the SP.

\subsection{Complexity evolution after a critical quench and dependence on the initial state}
\label{critical_quench}

Next we study the behavior of complexity for the critical quench, i.e. we assume that the post-quench magnetic field $h_{jf}=1$, in either of the two phases. Taking the limit $\omega_{jf} \rightarrow 0$, we see from eq. (\ref{complexity_2}) that for the critical quench the complexity grows with time as
\begin{equation}
	\mathcal{C}_j(t)\big|_{h_{jf}=1}=\frac{\omega_{ji}^2}{8}t^2~.
\end{equation}
Now for critical quenches starting from arbitrary values of the magnetic fields $h_{ji}$, the values of the $\omega_{ji}$s are not equal. Hence we observe different behavior of the SC in the two phases. In particular, in the BP, with an arbitrary initial magnetic field $h_{1i}$, the time evolution of complexity is different from that of in the SP with the initial magnetic field $h_{2i} (\neq h_{1i})$. 
In the following, we consider the time evolution of the SC at the critical quench for different initial magnetic fields, equal distances away from the critical value. 

\begin{figure}[h!]
	\centering
	\includegraphics[width=0.7\textwidth]{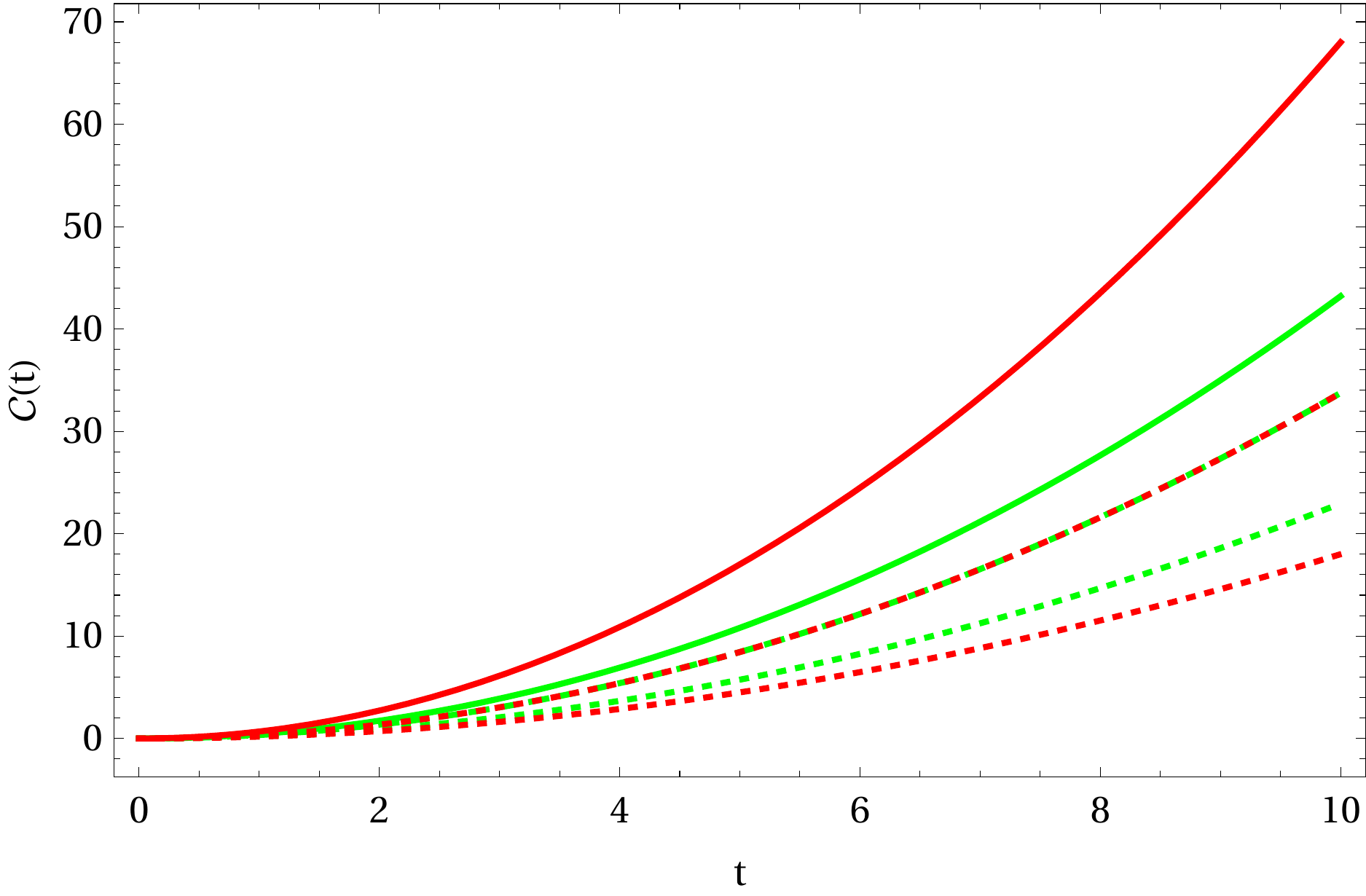}
	\caption{Evolution of $\mathcal{C}(t)$ in either phase of the system for critical quenches 
		with different initial states.
		Solid red and green curves are with $h_i=h_c\pm 0.8$, dotted red and green are with 
		$h_i=h_c \pm 0.3$. For a given $h_{1i}$ we can obtain a $h_{2i}$ for which the complexities 
		coincides. This is shown here with $h_{1i}=0.5$, for which we can fined $h_{2i}=1.486$ for the 
		crossover of complexities (dashed red and green curves). We have used $\gamma=0.1$.
	 }
	\label{fig:C_critical}
\end{figure}

Fig.~\ref{fig:C_critical} depicts the effect of changing the initial magnetic field on the critical quench. When the magnetic fields are far away from the critical point, the SC in the SP (solid red curve) grows faster than the one in the BP (solid green curve). However, when the initial magnetic fields take values closer to the critical point, the complexity in the BP (dotted green curve) grows faster than the SP (dotted red curve). Thus, one can notice that for a given $h_{1i}$, there is a particular value of $h_{2i}$ for which the time evolution of the complexities are identical in both phases. In particular, this happens when the two initial frequencies $\omega_{ji}$s are equal. In that case, we can conclude that the complexity evolution is continuous across the two phases.

An analytical formula for the relation between $h_{1i}$ and $h_{2i}$ can be derived from the expression for complexity in the critical quench derived above. This can be checked to be given by
\begin{equation}\label{hji}
h_{2i}=\frac{1}{2}\Big(\gamma+1 +\sqrt{(1-\gamma)(5-4 h_{1i}^2 - \gamma)}\Big)~.
\end{equation}
An example of such evolution is shown in Fig.~\ref{fig:C_critical} with $h_{1i}=0.5$ being the initial magnetic field in the BP. 
Utilizing the eq. (\ref{hji}), we can obtain the initial magnetic field in the SP $h_{2i}=1.486$ for the evolution to be identical in both phases after a critical quench.

Before concluding this section, we note that quadratic evolution of the complexity is a typical behavior shown by the complexity of the free particle \cite{Balasubramanian:2022tpr} \footnote{ A general argument was presented in \cite{Balasubramanian:2022tpr} explaining  the quadratic growth of the SC for free particles.}. In the thermodynamic limit, the Hamiltonian of the LMG model can be expressed as a harmonic oscillator with a vanishing frequency after the critical quench; hence, the system becomes a free particle. This provides a straightforward explanation of the quadratic growth of complexity after a critical quench.

\section{Time Evolution of spread entropy after a sudden quench}\label{SE}
In this section, we study another interesting quantity relevant in the context of the wave function spreading
 over the Krylov basis, namely the spread entropy defined in  eq. \eqref{entropy}. This quantity helps 
to understand the delocalization of the initial pre-quench state in the Krylov basis after a sudden quench. 

Utilizing the expressions for $\phi_{n}$s obtained in the eq. (\ref{phi_n}), we can express the spread entropy as follows
\begin{equation}
	\begin{split}
		S_{K}(t)=-|\phi_{0}(t)|^2 \bigg[\sum_{n}\mathcal{N}_n^2 \ln (\mathcal{N}_n^2) \mathcal{F}^n(t)
		+\ln (|\phi_{0}(t)|^2 )\sum_{n} \mathcal{N}_n^2 \mathcal{F}^n(t)\\
		+\ln (\mathcal{F}(t))\sum_{n} n\mathcal{N}_n^2 \mathcal{F}^n(t)\bigg]~.
	\end{split}\label{SKt}
\end{equation}
Here the numerical constants ($\mathcal{N}_n$) are given by
\begin{equation}
	\mathcal{N}_n=\frac{\Gamma(n+\frac{1}{2})}{\sqrt{\pi}\Gamma(n+1)}~.
\end{equation}
Using this  we can express the second and third terms of the \cref{SKt} in a compact form, so that the spread entropy can be written as,
\begin{equation}\label{K_entropy_ex}
	S_{K}(t)=-|\phi_{0}|^2 \bigg[\sum_{n=1}^{\infty}\mathcal{N}_n^2 \ln (\mathcal{N}_n^2) \mathcal{F}^n+
	\frac{\ln (|\phi_{0}|^2 )}{\sqrt{1-\mathcal{F}}}
	+\frac{\mathcal{F}\ln (\mathcal{F})}{2(1-\mathcal{F})^{3/2}}\bigg]~.
\end{equation}

\begin{figure}[h!]
	\centering
	\includegraphics[width=0.7\textwidth]{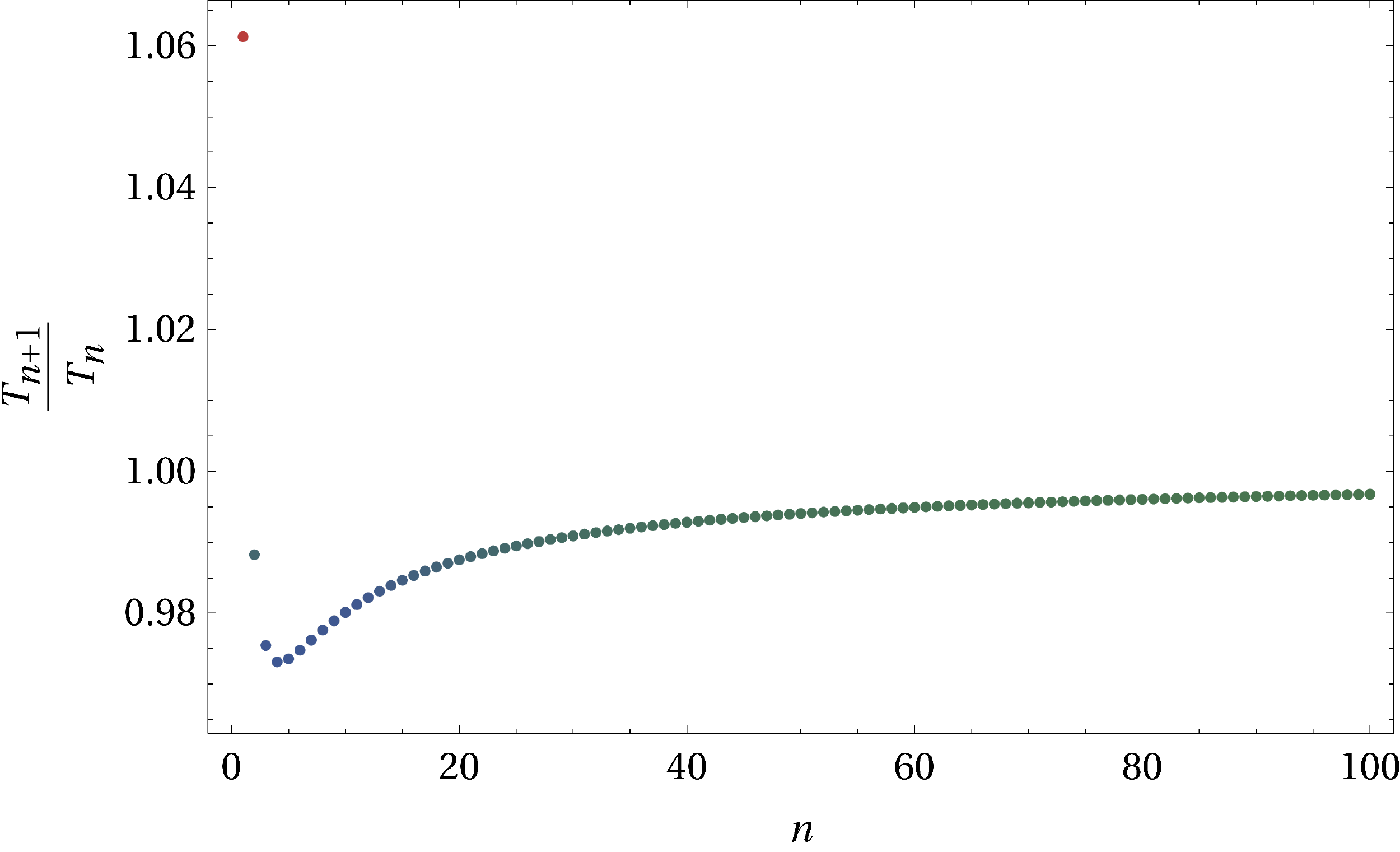}
	\caption{Plot of the ratio $\frac{\mathcal{N}_{n+1}^2 \ln (\mathcal{N}_{n+1}^2)}
		{\mathcal{N}_{n}^2 \ln (\mathcal{N}_{n}^2)}$. For higher values of $n$ the ratio approaches to a 
		limiting value around  $ \approx 0.995$}
	\label{fig:ratio}
\end{figure}

Since the first term in \cref{SKt} can not be written in  any further simpler form, we first study its behavior separately.
In Fig. \ref{fig:ratio}, we have plotted the ratio of the numerical coefficients of the successive terms in the first summation above. It can be observed that the ratio $\frac{T_{n+1}}{T_n}$  
(where $T_n=\mathcal{N}_{n}^2 \ln (\mathcal{N}_{n}^2)$)
 has a value greater than unity only for $n=1$, and, for higher values of $n$, it approaches a constant value less than unity.

On the other hand, we have already  shown that the function $\mathcal{F}(t)$ is always smaller than unity for
non-critical values of the magnetic field (see  figs.  \ref{fig:F_broken} and  \ref{fig:F_symmetric}). 
Hence, for the quench scenarios under consideration, the infinite sum can be replaced by the sum of a finite number of terms with a excellent  accuracy. This fact will help us study the evolution of spread entropy below.
In this section, we study the evolution of the spread entropy after a quantum quench in the BP and the SP, away from the criticality as well as at the critical point of the HO model. Special emphasis is given  on the behavior of the first term (the infinite sum) in \cref{K_entropy_ex}, which we call $T_1$ for convenience, i.e.
	$T_1=-|\phi_{0}|^2 \sum_{n=1}^{\infty}\mathcal{N}_n^2 \ln (\mathcal{N}_n^2) \mathcal{F}^n$.

\subsection{Post-quench evolution of spread entropy  for non-critical quenches}\label{quench_SE}
\begin{figure}[h!]
	\centering
	\includegraphics[width=0.7\textwidth]{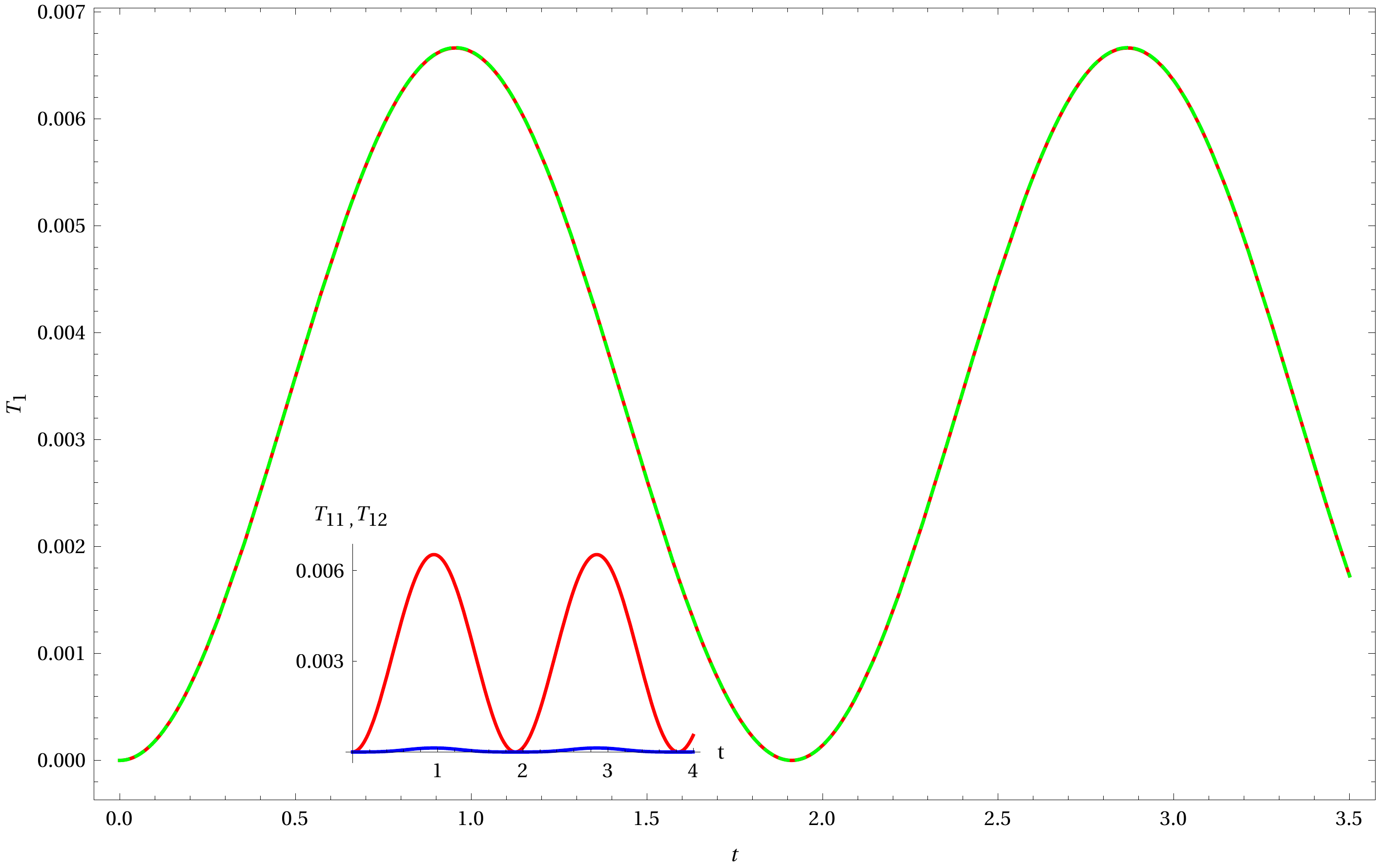}
	\caption{Plot of $T_1$, the first term in the expression for the spread entropy in the BP. 
		The red curve is sum of first two terms, while the green dashed curve 
		is sum of first 30 terms. Two plots are also identical implying that the sum converges rapidly.
		Here $h_f=0.5, h_i=0.1, \gamma_f=\gamma_i=0.1$. The inset shows the contribution of the $n=1$
		term (red curve, called $T_{11}$) and $n=2$th term (blue curve, called $T_{12}$)
		term in $T_1$.}
	\label{fig:1st_broken_ac}
\end{figure}

First we analyze the BP of the ground state when the HO is away from the critical point.
In Fig. \ref{fig:1st_broken_ac}, the red and the green curves show the time variation of $T_1$, when  the sum of the first two and thirty terms are considered, respectively. The identical character of these two curves for all 
values of time indicates that the contributions from the terms with higher values of $n$ are negligible. 
The inset of Fig. \ref{fig:1st_broken_ac} shows that the contribution of the  second term is in fact much smaller 
compared to the first term. Hence, we conclude that the infinite sum is rapidly convergent, and its value at any 
instant of time can be replaced by the sum of its first two terms when the system is in BP  and is 
sufficiently away from the critical point.

\begin{figure}[h!]
	\begin{minipage}[b]{0.53\linewidth}
		\centering
		\includegraphics[width=0.8\textwidth]{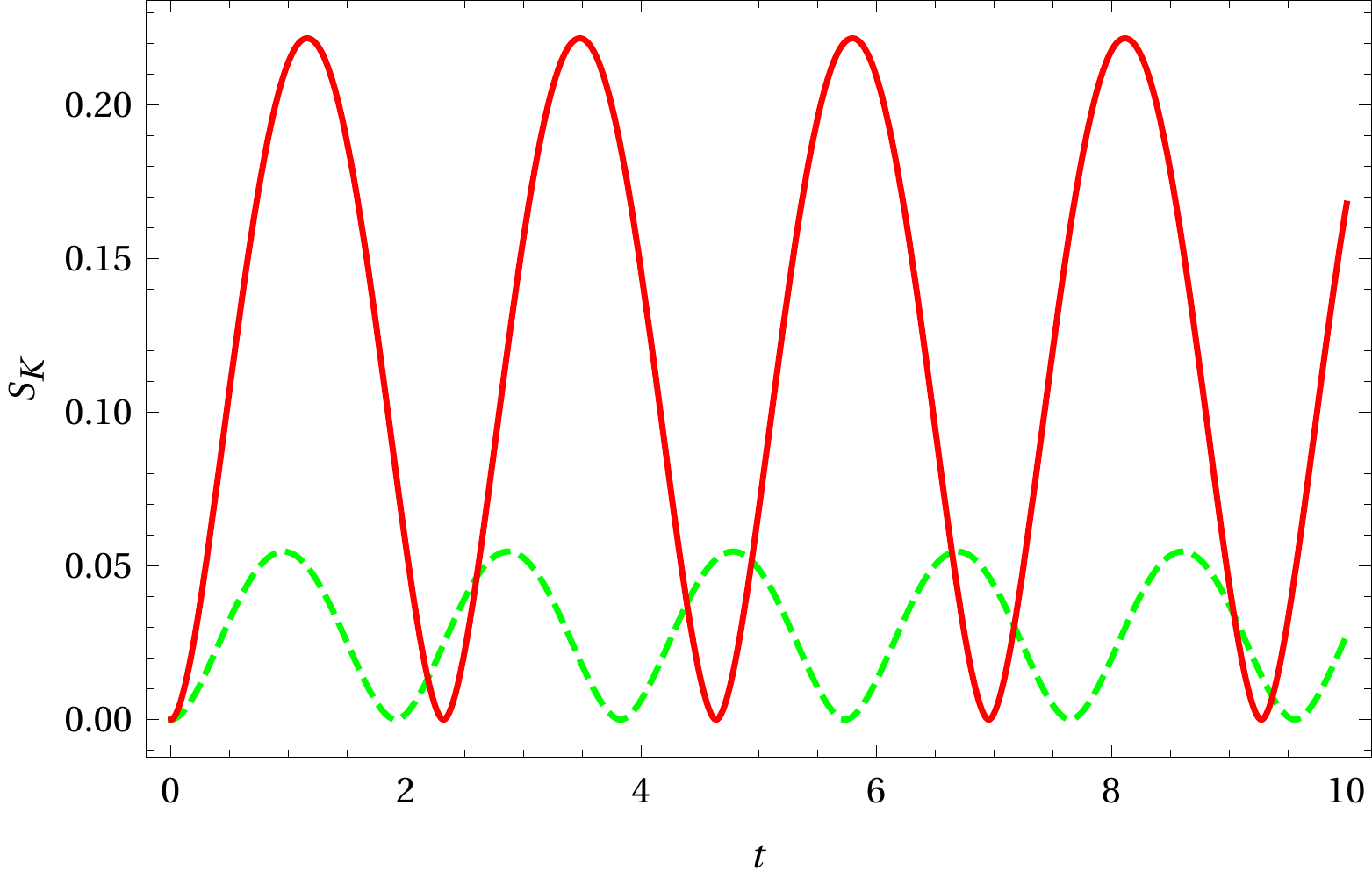}
		\caption{Time variation of the spread entropy in the BP when the magnetic field 
			is away from critical point. Here, $h_f=0.5$ (green) and $h_f=0.7$ (red), and $h_{1i}=0.1$.}
		\label{fig:kent_broken_ac}
	\end{minipage}
	\hspace{0.3 cm}
	\begin{minipage}[b]{0.53\linewidth}
		\centering
		\includegraphics[width=0.8\textwidth]{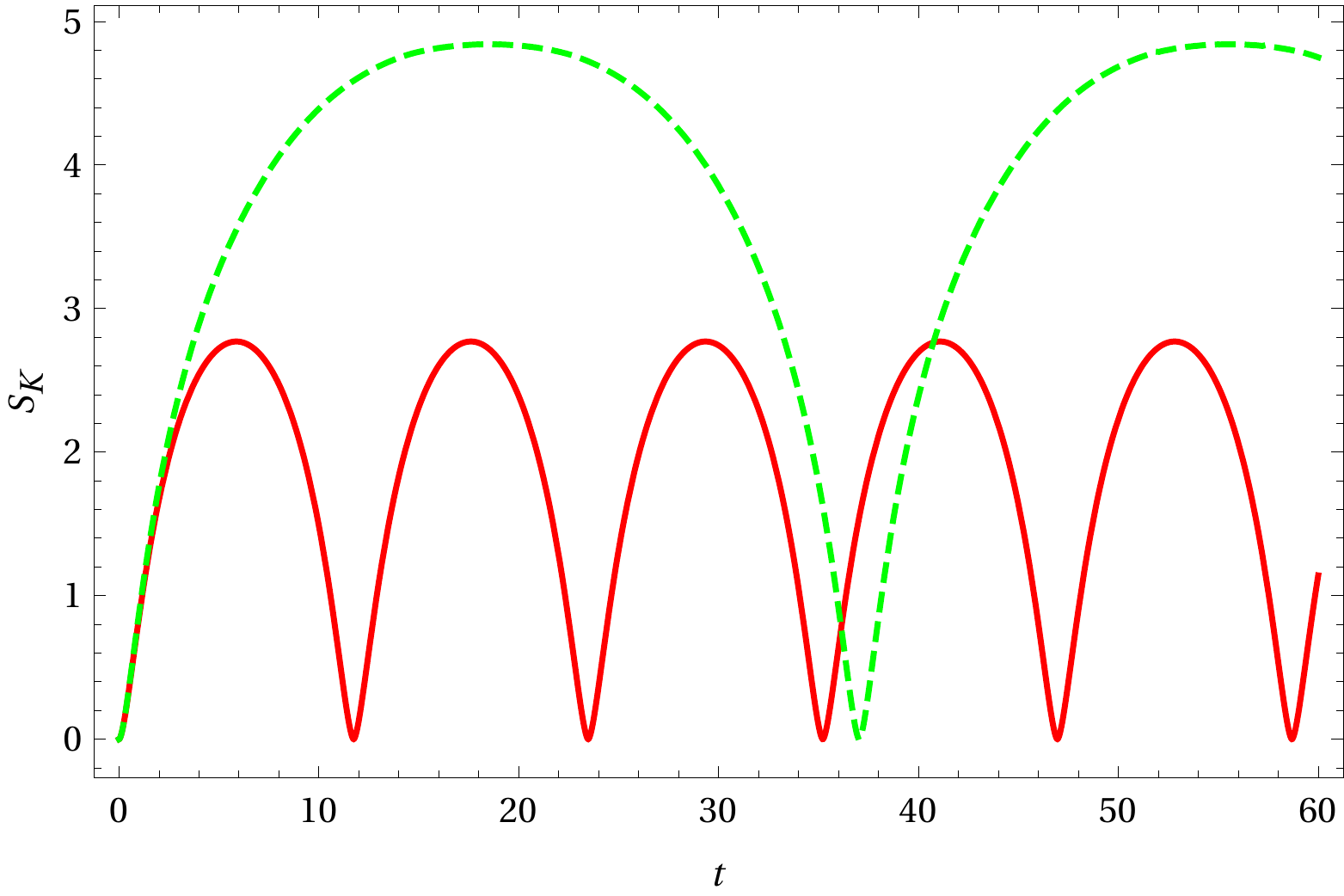}
		\caption{Time variation of the spread entropy in the BP when the magnetic field 
			is close to the criticality. Here, $h_f=0.99$ (red) and $h_f=0.999$ (green), and $h_{1i}=0.1$.}
		\label{fig:kent_broken_cc}
	\end{minipage}
\end{figure}

Now we consider different values of $h_f$ to probe the sensitivity of $S_K(t)$. In Fig. \ref{fig:kent_broken_ac}, we
have  plotted the spread entropy for two different values of the magnetic fields, $h_f=0.5$ and $h_f=0.7$ respectively, 
with $h_{1i}=0.1$. Here, we observe that the sum $T_1$ collects a considerable amount of contributions from the terms with higher values of $n$ as we take the magnetic field  closer to the critical value.\footnote{As we have discussed previously, this is of course true for the time evolution of the SC as well.} 
As an example, we set $h_f=0.7$, which is close to the critical magnetic field  compared to $h_f=0.5$. The time variation of the SC with $h_f=0.7$ is shown in Fig. \ref{fig:kent_broken_ac} (green dashed line) where only the first three terms have any meaningful contribution to $T_1$, in the sense that an equation analogous to \cref{N_eff} can be defined for the spread entropy. Furthermore, it can also be checked that the third term in \cref{K_entropy_ex} has the dominant contribution in $S_K(t)$, and the first term contributes the least.

Next we analyze the cases where the post-quench magnetic field value is very close to the critical point. It is observed that the number of terms with higher values of $n$, which contribute to $T_1$, increases rapidly. For $h_f=0.99$, the infinite sum $T_1$ saturates after summing up to $70$ terms, while for $h_f=0.999$ this number goes up to 300. The time evolution of the spread entropy for these two post-quench magnetic fields is plotted  in Fig. \ref{fig:kent_broken_cc}. 

For quenches in the SP, similar qualitative behavior of the spread entropy can be observed. However, the quantitative behavior between these two cases are  different. Specifically, the number of terms which contributes towards the sum $T_1$, such that, even  by adding the next term  $T_1$ does not increase
more than $\epsilon=0.001$ are not equal. \footnote{Once again, this statement is made by looking at the maxima
	of these time varying functions.}

For example, when we take $h_f=1.5$, at least terms with $n=1,2,3$ have a considerable amount of contribution in the sum. \footnote{Note that, in the BP, $T_1$ received the major contributions only from the first and the second term where the magnetic field was $h_{f}=h_c-0.5$. All these statements are  made by assuming 
	a cutoff of $\epsilon \approx 0.001$.} Analyzing the system close to the critical point i.e. $h_f=1.01$, we have to sum up to 175 terms to reach the saturation of $T_1$. Following the above discussion, we conclude that in the SP, the infinite sum $T_1$ converges slowly compared to the BP.

	The quantity defined in eq. \eqref{entropy} is the entropy  computed in the Krylov basis, and besides the SC, it 
	is also a measure of spreading of an initial state in the Krylov basis,
	as well as the participation of the higher order Krylov basis elements in the time evolution. 
	As we have mentioned before, 
	similar such quantities have been studied in the context of quantum quenches in interacting many-body quantum
	systems for  different choices of the basis. E.g., when computed in terms of the eigenstate of the 
	post-quench Hamiltonian, this is called  the diagonal entropy \cite{Barankov:2008qq}.  The entropy  is in fact 
	time-independent in such cases. On the other hand, it can also be computed in terms of the energy basis 
	of the pre-quench Hamiltonian, and in that case, it shows characteristics time evolution depending on the nature of the system under consideration and the initial state \cite{torres2014general}.

	The behavior of the spread entropy studied here for non-critical quenches is consistent with those 
	of the SC considered in the  previous section. In particular, we see that, though there is an infinite
	number in the  Krylov basis elements, for the purpose of numerical evolution 
	of the spread entropy (as well as the SC), we can, to good approximation, assume that the higher order Krylov basis
	elements  have not participated in the time evolution of the initial state. Therefore, the spread entropy 
	only oscillates with time. Of course, as we take the magnetic field closer to the critical point, 
	more and more basis elements start to participate in the spreading, and hence, the delocalization 
	of the initial state increases, thereby increasing  the entropy.

\subsection{Evolution of spread entropy after a  critical quench}\label{critical_SE}
We now study the evolution of spread entropy $S_{Kc}(t)$ when the sudden quench is critical. Considering the limit $\omega_{jf} \rightarrow 0$, we can express the three terms in \cref{K_entropy_ex} as follows (denoting them as $T_1, T_2$ and $T_3$ for convenience),
\begin{equation}\label{critical_terms}
	\begin{split}
		T_1(t)\big|_{\omega_{jf} \rightarrow 0}=-2\sum_{n=1}^{\infty} \mathcal{N}_n^2 \ln (\mathcal{N}_n^2) 
		\frac{\big(\omega_{ji}^2 t^2\big)^n}{\big(4+\omega_{ji}^2 t^2\big)^{n+1/2}}~,\\
		~~T_2(t)\big|_{\omega_{jf} \rightarrow 0}=-\frac{1}{2} \ln \bigg[\frac{4}{4+\omega_{ji}^2 t^2}\bigg]~,~~\\
	T_3(t)\big|_{\omega_{jf} \rightarrow 0}=-\frac{\omega_{ji}^2 t^2}{8}
		\ln \bigg[\frac{\omega_{ji}^2 t^2}{4+\omega_{ji}^2 t^2}\bigg]~.
	\end{split}
\end{equation}
We define a dimensionless parameter $x=\omega_{ji} t$ which controls the behavior of $T_1, T_2$ and $T_3$ in BP and SP. 
For small $x$, the spread entropy rises sharply at initial times, then after reaching a peak value, it starts to decrease until it hits a minimum, and finally, at late times, it diverges. The time at which the peak of $S_{Kc}(t)$ is reached can be determined by the frequency $\omega_{ji}$, and hence, by the value of the  initial magnetic field $h_{ji}$. For small values of $\omega_{ji}$, the peaks appear at later times, and the spread entropy diverges less rapidly at late times compared to the larger $\omega_{ji}$s. 

\begin{figure}[h!]
	\centering
	\includegraphics[width=0.7\textwidth]{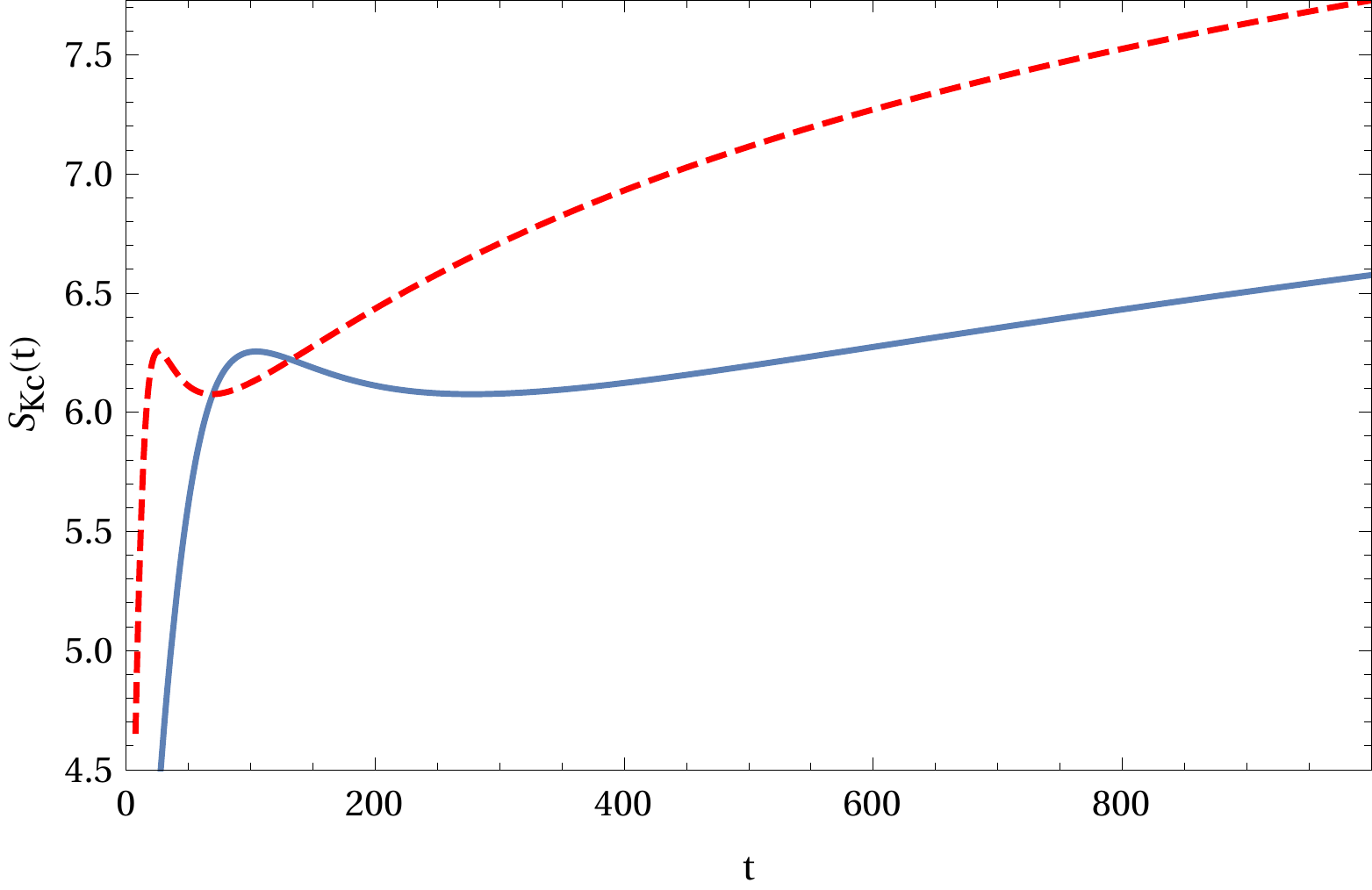}
	\caption{Time variation of spread entropy after a critical quench in the SP.
		The red curve is with $h_i=1.9$ ($\omega_{i} \approx 2.545$) and the blue curve
		is for $h_i=1.1$ ($\omega_{i} \approx 0.632$). }
	\label{fig:s_critical_symmetric}
\end{figure}

In Fig. \ref{fig:s_critical_symmetric}, we present the time evolution of the critical spread entropy ($S_{Kc}(t)$) for two different values of the magnetic fields in the SP. To explain the behavior of the plots, we analyze the early and late time characteristics separately for each of the terms in eq. (\ref{critical_terms}). 

For small values of $x$ (with fixed $\omega_{ji}$), the three terms in \cref{critical_terms} have the lowest order contributions as follows : $T_1 \approx \frac{\ln 2}{8}x^2$, $T_2\approx\frac{x^2}{8}$, and $T_3\approx -\frac{x^2}{8}\ln \big[\frac{x^2}{4}\big]$. From these expressions we can  observe that the third term has the dominant contribution 
in the growth of the spread entropy  at early times after the quench. On the other hand, at late times 
$T_1$, $T_2$ and $T_3$ show different behavior. The first term $T_1$  reaches a peak value and then decays to 
zero at late times, with the decay rate determined by the frequency $\omega_{ji}$. The second term grows 
continuously after the quench, and at late times, it diverges logarithmically, with the  rate of growth depending 
on the initial frequency. Finally, the third term, after the initial growth, saturates to a constant value of $1/2$ 
at the limit $x \rightarrow \infty$ irrespective of the frequency $\omega_{ji}$.

When we take  the initial magnetic field to be  close to the final magnetic field (which is equal to unity here), the critical spread entropy shows a slow increase up to a peak value and then falls off slowly compared to the case when the initial magnetic field is away from the final value. The rate of growth at late times is also slow in this case.
In both the cases shown in Fig. \ref{fig:s_critical_symmetric}, we have taken the sum of the first 900 terms to compute $T_1$, which provides an excellent convergence for small values of $t$. However, for large $t$ this is slightly inaccurate. It can be checked that the contribution of $T_1$ towards $S_{Kc}(t)$ is much smaller compared to the second term. At  late times,  $T_2$ shows logarithmic divergence, which becomes the dominant contribution in $S_{Kc}(t)$. Thus, the restriction in the accurate computation of $T_1$ does not affect the main behavior of the spread entropy as  presented here. 

The divergence of the spread entropy at late times for critical quenches is also consistent with the behavior 
of the SC for such quenches.  In this case, the initial state delocalize  over the entire Krylov basis, and hence
the entropy diverges at late times.

Before moving on to the next section we notice that similar behavior of the critical spread entropy as shown in Fig. \ref{fig:s_critical_symmetric} can also be observed for quench in the BP of the HO system. 
Furthermore, for a given magnetic field in the BP, we can obtain a particular value of the magnetic field where the evolution of critical spread entropies are identical in two phases. Note that a similar behavior was observed in \cref{spread_complexity} for SC.

\section{Time evolution of the spread complexity in the LMG model after a sudden quench }
\label{SC_LMG_Exact}
In section \ref{SC_LMG} we have studied the time evolution of the SC after a sudden quantum
quench in a HO model related to the LMG model at the thermodynamic limit.
In this section we briefly study the time evolution of the SC in a  quenched  LMG model. 
We have briefly reviewed the main properties of the LMG model in the thermodynamic limit 
at the beginning of section \ref{Therm_LMG}. For more details of this model,
we  refer to \cite{PhysRevLett.99.050402,PhysRevE.78.021106, PhysRevB.71.224420}.

The primary 
difference between the scenarios considered here and in section \ref{SC_LMG} is the following.
In this section we assume that the parameters of the LMG model in \cref{LMGHamiltonian} 
are quenched to a new set of values, and the SC of the resulting system is studied at the thermodynamic limit.
On the other hand, in section \cref{quench} we quenched a HO model whose frequency formally coincides with 
those of the LMG model written at the thermodynamic limit (see \cref{BP-Hamiltonian}, \cref{SP-Hamiltonian} above).
Mathematically, the  difference between the two models, as we shall see in 
the sequel, is the relation between   the creation and annihilation operators before and after the 
quench. The basic steps of the calculation of SC are same as before. Therefore, here we only briefly outline
the computation of the SC for brevity.

The  quench protocol we consider here is slightly different from the cases considered in \cref{SC_LMG}, in particular
the protocol in this section is a relatively restricted one in the BP.  When the system is in the SP of the ground state we change the 
$h_i$ and $\gamma_i$ suddenly to a new  values $h_f$ and $\gamma_f$ respectively.  However in the BP,
the magnetic field before and after the quench remains the same ($h_i=h_f$) where only the anisotropy parameters
are different.  For a general quench in the BP  (where both the magnetic field and the anisotropy
parameter are changed), it is actually relatively difficult to obtain the relationship between the 
creation and annihilation operators before and after the quench. This issue is discussed towards  the end of this
section.

The LMG Hamiltonian of \cref{LMGHamiltonian} can be written in terms of a set of bosonic operators $\beta$ and $\beta^\dagger$ after the HP transformation as \cite{PhysRevB.71.224420}, 
\begin{equation}
	\mathcal{H}=\Delta \beta_i^{\dagger}\beta_i+\Gamma (\beta_i^{\dagger 2}+\beta_i^2)~,~~i=1,2~,
\end{equation}
where
\begin{equation}
	\Delta=2+2Bm-3m^2-\gamma~,~~~\text{and}~~~\Gamma=\frac{\gamma-m^2}{2}~.
\end{equation}
In the BP and SP $m$ is equal to $h$ and 1 respectively. The subscript in $\beta_i$ represents which one of the two phases is under
 consideration. To diagonalize this Hamiltonian we perform the following Bogoliubov transformation
\begin{align}
	\label{Bgogliubov_exact}
	 \beta_i=	U_i \alpha_i+ V_i \alpha_i^{\dagger}~,~
   \beta_i^\dagger=U_i \alpha_i^\dagger+ V_i \alpha_i,
\end{align}
where $U_i=\cosh (\Theta_i/2)$ and $V_i=\sinh (\Theta_i/2)$. The final form of the Hamiltonian after this 
transformation is given by
\begin{equation}
	\mathcal{H}=\omega_{i}\alpha_{i}^{\dagger}\alpha_i~.
\end{equation} 
The frequency of the resulting oscillator $\omega_{i}$ in the two phases have been already  provided in
\cref{BP-Hamiltonian,SP-Hamiltonian}.  The angles $\Theta_i$ are also different in the two phases
and are given by
\begin{align}
	\tanh \Theta_1=\frac{h^2-\gamma}{2-h^2-\gamma}~~ \text{for}~~h<1~,\\
	\tanh \Theta_2=\frac{1-\gamma}{2h-1-\gamma}~~ \text{for}~~h>1~.
\end{align}
 To obtain the time evolved state after the quench in either of the two phases, we need to find out relation between the creation and the annihilation operators before and after the quench. We call the 
 bosonic operators after the quench as $\tilde{\alpha}_i$ and $\tilde{\alpha}_i^\dagger$ respectively.
Using the transformation relations in \cref{Bgogliubov_exact} we can obtain the relation between these 
two sets of operator to be
\begin{equation}
	\label{Bogoliubov_quench}
	\tilde{\alpha}_i=\mathcal{U}_i\alpha_{i}-\mathcal{V}_i\alpha_{i}^\dagger~~,~~
	\tilde{\alpha}_i^\dagger=-\mathcal{V}_i\alpha_{i}+\mathcal{U}_i\alpha_{i}^\dagger~,
\end{equation} 
where $\mathcal{U}_i=\cosh\Big[\frac{\tilde{\Theta}_i-\Theta_i}{2}\Big]$, and $\mathcal{V}_i=\sinh\Big[\frac{\tilde{\Theta}_i-\Theta_i}{2}\Big]$. Since these relations are analogous to 
\cref{Bogoliubov}, the rest of the calculations follow the same steps as before.
The expression for the auto-correlation function in this case is given by
\begin{equation}
	\mathcal{S}(t)=\Big[\cos (\tilde{\omega}_i t)- 
	i \cosh (\tilde{\Theta}_i-\Theta_i)\sin (\tilde{\omega}_i t)\Big]^{-1/2}~,
\end{equation}
where $\tilde{\omega}_i$ is the frequency after the quench. Using these $\mathcal{S}(t)$ we calculate the Lanczos coefficients and the expansion coefficients
$\phi_n$ corresponding to the time evolved wavefunction in the Krylov basis. The expressions for $a_n$ and $b_n$
are the same as those given in \cref{Lanczos}, with $\mathcal{U}_i$ and $\mathcal{V}_i$ given just 
below \cref{Bogoliubov_quench}. On the other hand  $\phi_n$s are given by
\begin{equation}\label{phi_n_exact}
	\phi_{n}(t)=\mathcal{N}_n \phi_0 (t)\mathcal{G}(t)^n~, 
	~~ \mathcal{G}(t)=\frac{\cosh \chi \sin (\tilde{\omega}_i t) \sinh \chi}
	{\cosh (2\chi) \sin (\tilde{\omega}_i t) - i \cos (\tilde{\omega}_i t) }~.
\end{equation}
Here, as usual $\phi_0=\mathcal{S}(t)^*$, and we have denoted $\chi=(\tilde{\Theta}_i-\Theta_i)/2$.
We provide the first few vales of the numerical constants: $\mathcal{N}_n$ : $\mathcal{N}_1=1, 
\mathcal{N}_2=\sqrt{2}, \mathcal{N}_3=\sqrt{\frac{3}{2}}, \mathcal{N}_4=\sqrt{\frac{10}{9}}, \cdots$.
Using these expressions we calculate the SC to be 
\begin{equation}
	\mathcal{C}(t)=2|\phi_1(t)|^2\sum_{n=0}^{\infty} 2^{2n}\frac{(n+1) (2n+1)!!}{(2n+2)!!}\mathcal{F}^{n}(t)
	=\frac{|\phi_1(t)|^2 }{\big(1-4\mathcal{F}(t)\big)^{3/2}}~,\text{where}~~
	\mathcal{F}(t)=|\mathcal{G}(t)|^2~.
\end{equation}
In this case, the expression of $\mathcal{F}(t)$ is given by
\begin{align}
	\mathcal{F}(t)=\frac{\cosh ^2\chi \sin ^2(\tilde{\omega}_i t) \sinh^2 \chi}
	{\cosh ^2(2\chi) \sin ^2(\tilde{\omega}_i t) + \cos ^2(\tilde{\omega}_i t) }~.\label{ct}
\end{align}
This expression for the SC can be simplified further by using the explicit form of $|\phi_1(t)|^2$ as 
\begin{align}
	\label{complexity_exact}
	\mathcal{C}(t)=\frac{1}{2}\sinh ^2 (2\chi) \sin ^2(\tilde{\omega}_i t)~.
\end{align}
Comparing \cref{complexity_exact,complexity_2}, we observe that the time dependence of $\mathcal{C}(t)$ on the post-quench frequency is 
the same in both the models, however the exact dependence of the oscillation magnitudes on the parameters are different. The plots for the time dependence of the complexity are thus similar to the previous case, and 
show oscillatory behaviour with time, and therefore, we do not provide them here for brevity. 

To quantify  the time evolution of the initial state in the Krylov after a sudden quench in the LMG
model  we can find out the $N_{eff}$ introduced in eq. \eqref{N_eff}. However, the behaviour of $N_{eff}$ is entirely 
similar to those  of the HO model considered in section \ref{SC_LMG} and  shown in Fig. \ref{fig:eff_symmetric}
for the SP of the HO.
\footnote{For the BP of the LMG model, as we have discussed previously, the quench setup is more restricted,
so that  we can not directly compare the $N_{eff}$ for the BP with that of the SP. }
Therefore, we have not shown these plots  here for brevity.

	The periodic time dependence of the SC observed after a quench in the LMG model in the 
	thermodynamic limit, as well as in the auxiliary HO model studied in section 3, can be related 
	 to the fact that both of these models are integrable in nature. Mathematically, the ratio of the modulus squared
	of two wavefunctions $\phi_n$ is smaller than unity, so that  the sum in the expression of the SC can be performed exactly.
	This is due to the fact that the models we are considering are essentially single-mode harmonic oscillators,
	and hence, for a given magnetic field,
	the time evolved state has most of the support over up to a certain element of the Krylov
	basis. 
	Here we also mention that similar behaviour of KC has also been observed 
	in field theories compactified on a sphere  \cite{Avdoshkin:2022xuw}.

Before concluding this section, we discuss here why the quench protocol in the BP of the LMG model is restricted 
to the special case $h_f=h_i$. In the BP, before applying the HP transformations, we need to rotate the spin operators appearing in \cref{LMGHamiltonian} which align the $z$ axis along the direction of the semi-classical magnetization. In the BP, the rotation angle depends on the external magnetic field \cite{PhysRevB.71.224420}
(in SP, the angle of rotation is zero). Hence, a quench scenario in this phase involving two different magnetic fields will correspond to two different rotation angles, which differentiates the spin operators after the rotation for pre-quench and post-quench scenarios. As a consequence, the relations between the bosonic creation and annihilation operators before and after the quench are difficult to obtain (this relation is given in \cref{Bogoliubov_quench} for the particular case we have considered).

\section{Conclusions}\label{conclu}
In this paper, we have used the SC as a probe of equilibrium QPT  when a system is quenched towards the quantum critical point,
by considering the paradigmatic LMG model with spins having infinite range interactions as an example. 
We have defined a new quantity $N_{eff}$, which measures the total number of elements of the Krylov basis that contributes to the summation of SC up to a pre-defined cutoff. 
We have shown that this quantity grows as the post-quench magnetic field is taken closer to the critical point 
in both phases of the LMG model. 
We argue that this measure of effective basis can also distinguish between two phases of the LMG model since the exact nature of growth in this quantity is different in the two phases. 
When the sudden quench is a critical quench, we have shown that the SC grows quadratically with time. This behavior of the SC can be anticipated from the fact that the system for a  critical quench behaves like a free particle. This is in accordance with the arguments given in \cite{Balasubramanian:2022tpr}.

A similar analysis for spread entropy reveals an oscillatory behavior when the final value of the magnetic field is away from the critical point in both phases. On the other hand, near criticality, the spread entropy still exhibits an oscillatory behavior but with a larger amplitude and time period. However, in the SP, the infinite sum in the spread entropy converges slowly compared to the BP. When the sudden quench is a critical one, the spread entropy
diverges logarithmically at late times (whereas the SC diverges quadratically with time).

In this paper, we have taken the first steps towards understanding the ground state QPTs using quantities that characterize the spreading of an initial state
	before the quench with time in terms of the Krylov basis. In this context, it is worthwhile to compare and contrast this paper with other works on studying QPTs using different measures of complexity. Starting with the works of \cite{Liu:2019aji, Ali:2018aon}, where the NC was used to probe topological phase transitions, quite a number of studies have appeared along these lines, see e.g.,  \cite{Pal:2022ptv, Jaiswal:2020nzq, Jaiswal:2021tnt, Camilo:2020gdf, Sood:2021cuz, Huang:2021xjl}. However, exploring the zero-temperature QPTs using the recently proposed notion of SC has one significant advantage is that  this measure of complexity does not depend on the choice of cost-functional - 
	an essential feature (as well as a drawback in many cases) of Nielsen-type complexities. Consequently, in some sense, SC can be thought to  provide 
	a robust characterization of QPTs
	using information theoretic quantities. Furthermore, in refs. \cite{Caputa:2022eye, Caputa:2022yju}, the SC was used to study topological phase transition in SSH and Kitaev models, respectively, both in and out-of-equilibrium scenarios, and the utility of SC as a marker of phase transition was established. In our work, though, we have mainly focused on the zero temperature 
	QPTs in infinite range interacting LMG model, we briefly consider quenches in the SSH model showing the topological 
	phase transition in the Appendix \ref{SSH} below. There, we have also highlighted the main differences between our
	results and those of ref \cite{Caputa:2022eye}.

Before concluding we note that for the infinite range interacting LMG model we have chosen, 
it is possible to write down the analytical expressions for the Lanczos coefficients explicitly, and the summation
in the SC converges  towards a finite value. However, this is not guaranteed to be the case
for more complicated models, and an appropriate regularization procedure might be needed. Consequently, many of the features that appeared in our case might be altered, and we leave this kind of analysis for future works.

\section*{Acknowledgements}
We sincerely thank Diptarka Das, Suchetan Das, and Tapobrata Sarkar for discussions and their valuable comments on a draft version of this manuscript. We also acknowledge MHRD, India, for providing the research Fellowship. BD would also like to acknowledge the support provided by the Max Planck Partner Group grant MAXPLA/PHY/2018577. We also remember fond memories of the Hall-X corridors'' where the main idea of this article was generated and many of the discussions took place. We thank our anonymous referees for their constructive comments and criticisms, which helped us to improve the draft version of this manuscript.

\appendix
\section{Spread complexity in quantum quench of the SSH model}\label{SSH}

In this appendix we briefly describe the time evolution of the SC after sudden quenches in the SSH model. Here the state before the quench is taken to be the first state of the Krylov basis .\footnote{This is the lowest weight state of a $su(2)$ representation in the context of the SSH model.}
In this case, only 
finite number of  Lanczos coefficients $a_n$ and $b_n$ are non-zero, and 
as a result only finite numbers of terms contribute in the sum  eq. (\ref{spread_complexity})
in the definition of the SC. This is in contrast with the LMG model 
considered in the main text where there are infinite number of non-zero
Lanczos coefficients $a_n$ and $b_n$ leading to the fact that the SC is  an infinite sum of non-zero
terms. Due to  this  difference in the   behavior  of the Lanczos coefficients,  SC shows
different characteristics during time evolution in the two cases.

We  consider a simplified  version of the SSH model given by the following 
Hamiltonian, written in terms of two-flavored fermion  creation and annihilation operators
$(\alpha_{i},\alpha_{i}^{\dagger})$ and  $(\beta_i,\beta_i^{\dagger})$ as,
\begin{equation}\label{ham}
	\mathcal{H}=q_1\sum_k \Big(\alpha_{i}^{\dagger}\beta_i+h.c.\Big)-
	q_2\sum_k \Big(\beta_{i}^{\dagger}\alpha_{i+1}+h.c.\Big)~.
\end{equation}
Here, we assume that the two parameters $q_1$ and $q_2$ take only real positive
values $q_1\geq0$ and $q_2\geq 0$. The system exhibits two different phases depending
on the relative values of these two parameters. For $q_1>q_2$, the system is 
in the non-topological phase and for $q_1<q_2$, the system is in the topological
phase. These two phases are separated by a critical point at   $q_1=q_1$.

In  the momentum space, the Hamiltonian of the SSH model in \cref{ham} can be written as
\cite{Miyaji:2014mca}
\begin{equation}
	\mathcal{H}=\sum_k \Big[2r_3 J_0^{(k)}+ir_1\Big(J_{+}^{(k)}-J_{-}^{(k)}\Big)\Big]~.
\end{equation}
Here the coefficients $r_i$ are related to the original parameters  $q_i$s by the following relations 
\begin{equation}
	r_1=q_1-q_2 \cos k~, \quad \text{and}  \quad r_3=q_2 \sin k~,
\end{equation}
and $J_i^{(k)}$ are the generators of the $su(2)$ algebra which satisfy the usual commutation
relations for each momentum modes. In the momentum space, the SSH Hamiltonian we have considered
is thus an element of the $su(2)$ algebra. Hence the ground state of the system is actually a
direct product of  spin coherent states corresponding to each modes \cite{Caputa:2022eye}.
For future convenience we define an angle $\theta_k$ as
\begin{equation}
	\sin \theta_k=\frac{|r_1|}{r}~, ~~ \cos \theta_k=\frac{r_3}{r}~, ~~\text{with}~~ 
	r=\sqrt{r_1^2+r_3^2}=\sqrt{q_1^2+q_2^2-2q_1 q_2 \cos k}~.
\end{equation}

To study the time evolution of the SC in the SSH model we consider the 
following quench protocol. We assume that before the quench the system is 
prepared in state
$\Ket{\frac{1}{2},-\frac{1}{2}}$, which is  a direct product of the states of the form
$\Ket{\frac{1}{2},-\frac{1}{2}}_k$. Each of these states $\Ket{\frac{1}{2},-\frac{1}{2}}_k$ 
is annihilated by the operator $J_-^{(k)}$. 
This state is the first state of the Krylov basis  i.e. $\ket{K_0}=\Ket{\frac{1}{2},-\frac{1}{2}}$.
At $t=0$ we instantaneously change the parameters $q_{1}^i,q_{2}^i$ before the quench to
a new set of values $q_{1}^f,q_{2}^f$. The subsequent time evolution of the system is 
generated by the new Hamiltonian $\mathcal{H}_f$ corresponding to these new set of parameters.

Before delving into the calculations of the SC in such a quench protocol 
we notice the following important points. 
The SC in a quantum quench protocol of the SSH model we are considering
has been studied recently in \cite{Caputa:2022eye}.  However in that work, the state of the 
system before the quench is assumed to be the ground state of the initial Hamiltonian, 
so that the first element of the Krylov basis is not the same as the state before the 
quench. On the other hand we have assumed that the initial state is the lowest weight
state of the $SU(2)$ group which is not the ground state of the system. 
After computing the SC we will compare  these two  results.

Let us first study the time evolved state after the quench, given for each 
momentum mode as
\begin{equation}
	\ket{\Psi_k(t)}=\exp \bigg[-it\Big(2r_{3f} J_0^{(k)}+ir_{1f}
	\big(J_{+}^{(k)}-J_{-}^{(k)}\big)\Big)\bigg]\Ket{\frac{1}{2},-\frac{1}{2}}_k~.
\end{equation}
Using the well known decomposition formula for the $SU(2)$ group elements 
\footnote{See \cite{Ban:93} for derivations of such formulas.} we can easily 
see that the time evolved state is a linear combination of only two basis elements,
namely\footnote{There can be a normalization factor difference from these basis states.},
$\Ket{\frac{1}{2},-\frac{1}{2}}$ and $\Ket{\frac{1}{2},\frac{1}{2}}$.
Essentially, theses are the  elements of the Krylov basis. Thus it can be 
anticipated that  only finite numbers  of Lanczos coefficients can have  nonzero 
values. This fact also makes the calculation of the SC much simpler.

Using this time evolved state we can calculate auto-correlation function $\mathcal{S}(t)$
as  
\begin{equation}\label{1}
	\mathcal{S}_{SSH}(t)=\braket{\Psi_f(t) | \psi(t=0)}=\cos (r_f t)-i\cos \theta_f \sin(r_f t)~.
\end{equation}
We notice that the expression in \cref{1}
is different from the one obtained in \cite{Caputa:2022eye}. In particular, $\mathcal{S}_{SSH}(t)$
given above does not depend on the  parameters of the initial Hamiltonian. However this is expected
since we have assumed the system to be prepared in the state $\Ket{\frac{1}{2},-\frac{1}{2}}$,
which is not the ground state of the system, and hence does not depend on the parameters of 
the initial Hamiltonian. 

Using the Lanczos algorithm we can calculate the non-zero Lanczos coefficients in this case to be
\begin{equation}
	a_0=-r_f \cos \theta_f~,~~a_1=r_f \cos \theta_f~,~~\text{and}~~b_1=r_f |\sin \theta_f |~.
\end{equation}
Next, we express the expansion coefficients $\phi_0(t)$ and $\phi_1(t)$ of the time evolved state in the Krylov
basis:
\begin{equation}
	\phi_{0}(t)=\cos (r_f t)+i\cos \theta_f \sin(r_f t)~, ~~~\text{and}~~~ 
	\phi_{1}(t)=-i|\sin\theta_f| \sin(r_f t)~.
\end{equation}
Using these, we obtain the expression for the SC of a single momentum mode, written in terms of the 
parameters of the SSH model as
\begin{equation}
	\mathcal{C}_k(t)=\frac{\big(q_1^f-q_2^f \cos k\big)^2}{(q_1^f)^2+(q_2^f)^2-2q_1^f q_2^f \cos k}
	\sin ^2 \Big(\Big[(q_1^f)^2+(q_2^f)^2-2q_1^f q_2^f \cos k\Big]t\Big)~.
\end{equation}

In the continuum limit, the total complexity is obtained by integrating  over the entire range of $k$
and multiplying by a factor of 2, incorporating the negative $k$ modes.

\begin{figure}[h!]
	\centering
	\includegraphics[width=0.75\textwidth]{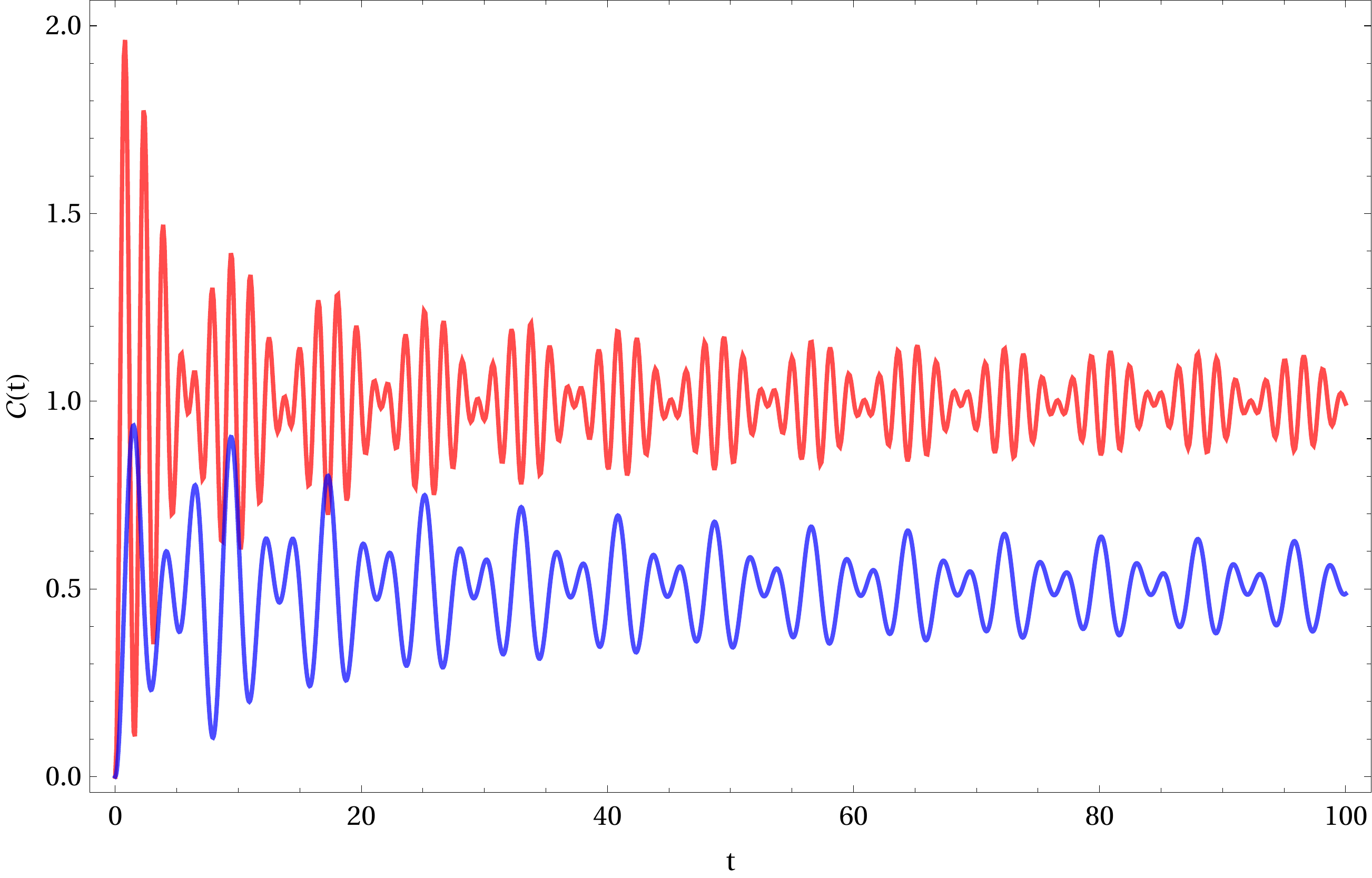}
	\caption{Time evolution of spread entropy after quench in the SSH model.
		The red curve is with $q_1^f=2, q_2^f=0.2$, and the blue curve is with $q_1^f=0.2, q_2^f=1$. }
	\label{fig:s_ssh_complexity}
\end{figure}

In Fig. \ref{fig:s_ssh_complexity}, we have shown the time evolution of the SC
after quantum quenches in the SSH model for two different final states computed by numerically integrating the 
expression for the SC for each mode obtained above. The red plot shows the time evolution when the
post-quench state is in the non-topological phase, while the blue curve indicates the evolution when the
final state is in the topological phase. The general behavior of the SC
is oscillatory with decaying magnitude, and after a sufficient amount of time after the quench, we see that
the SC of the non-topological phase is always larger than the topological phase.  
If we now perform two different quenches of the parameters of the SSH model, wherein the first  
case, the final state is in the non-topological phase, and  in the second case, the final state 
is in the topological phase, we may then distinguish these two phases by utilizing the evolution curves for the SC under these different quench protocols. 

We also notice that interchanging the values of $q_1^f$ with $q_2^f$, the magnitudes of $\mathcal{C}_k(t)$ becomes different, however, the time period of oscillations remains unchanged.
This  behavior of the SC we notice here is different  compared to the one obtained in \cite{Caputa:2022eye},
although their general oscillatory nature is the same.
If we consider two quench protocols with the final states in two different phases where the two parameters
are being just interchanged,
the SC then oscillates around two different values in our case.
On the other hand, according to the results of  \cite{Caputa:2022eye},
the SC oscillates around a common value irrespective of the final phase of the system.
It would be interesting 
to quantify this difference further.

\bibliographystyle{JHEP}
\bibliography{refs}

\end{document}